\documentclass{aa}
\usepackage{epsfig}
\usepackage{rotating}
\usepackage{txfonts}

\newcommand{\gesssim}{\mathrel{\hbox{\rlap{\hbox{\lower4pt\hbox{$\sim$}}}\hbox{$>$}}}}

\newcommand{\teff}{$T_{\rm eff}$}

\begin{document}

\title{Old open clusters as key tracers of Galactic chemical evolution.  I. Fe
abundances in NGC~2660, NGC~3960, and Berkeley 32 \thanks{Based on observations
collected at ESO telescopes under programs 072.D-0550 and 074.D-0571}}

   \subtitle{}

   \author{P. Sestito\inst{1,2,3} \and A. Bragaglia\inst{3} \and S.
Randich\inst{2} \and E. Carretta\inst{3} \and L. Prisinzano\inst{1} \and M.
Tosi\inst{3}}

   \offprints{P. Sestito, email:sestito@arcetri.astro.it}

\institute{INAF-Osservatorio Astronomico ``G.S. Vaiana'' di Palermo, Piazza del Parlamento 1, I-90134 Palermo, Italy
\and
INAF-Osservatorio Astrofisico di Arcetri, Largo E.~Fermi 5,
             I-50125 Firenze, Italy
\and
INAF-Osservatorio Astronomico di Bologna, Via C. Ranzani 1,
            I-40127 Bologna, Italy}

\titlerunning{Chemical abundances in open clusters}
\date{Received Date: Accepted Date}

  \abstract
   {}
{We obtained high-resolution UVES/FLAMES observations of
a sample of nine
old open clusters spanning a wide range
of ages and Galactocentric radii. The goal of
the project is to
investigate the radial metallicity
gradient in the disk, as well as the abundance
of key elements ($\alpha$ and Fe-peak elements).
In this paper we present the
results for the metallicity of three clusters: NGC~2660
(age $\sim$ 1 Gyr, Galactocentric distance of 8.68 kpc),
NGC~3960 ($\sim$1 Gyr, 7.80 kpc), and Be~32 ($\sim$6--7 Gyr, 11.30 kpc).
For Be~32 and NGC~2660, our study provides the first metallicity determination
based on high-resolution spectra.}
{We performed equivalent width analysis  with the spectral code MOOG,
which 
allows us to define a metallicity scale and build a homogeneous sample.}
{We find that NGC~3960 and NGC~2660 have a metallicity 
that is very close to solar
([Fe/H]=+0.02
and +0.04, respectively), while the older Be~32 turns out to have
[Fe/H]=$-$0.29.}
{}

\keywords{ Stars: abundances --
           Stars: Evolution --
           Galaxy:disk --
           Open Clusters and Associations: Individual: NGC~2660, NGC~3960,
	   Berkeley 32 }
\maketitle
\section{Introduction}\label{intro}

The investigation of chemical abundances in stars is fundamental for 
comprehending  the Galaxy formation mechanisms and the subsequent evolution.
In recent years, several observational and theoretical studies have been
carried out, aimed at understanding various features, 
such as the star formation
history  in the disk and its Galactocentric distribution, the gas distribution
and stellar density, the metallicity distribution ([Fe/H]) in the disk, as well
as the abundances of other elements relative to Fe (e.g. Edvardsson et
al.~\cite{edvar}; Friel
et al.~\cite{friel03}; Carretta et al.~\cite{carretta04}, \cite{carretta05};
Yong, Carney, \& de Almeida \cite{yong05}). 
Although many observational characteristics of the Milky Way are 
reproduced well by 
the appropriate theoretical models (Lacey \& Fall \cite{lacey}; Tosi
\cite{tosi88}; Giovagnoli \& Tosi \cite{giova};  Chiappini, Matteucci, \&
Gratton \cite{chiappini97}; Boissier \& Prantzos \cite{boissier}), several
problems persist.

One of these problems 
concerns the radial metallicity gradient in the disk, i.e. the
distribution of chemical elements with Galactocentric distance ($R_{\rm {gc}}$)
and its evolution with age. The radial metallicity gradient 
and its temporal evolution are among the most
critical constraints on Galactic chemical evolution models, since
predictions of its slope and slope variations with time depend mainly on the
relative timescales of the gas consumption (i.e. star formation rate) and gas
accretion (i.e. infall rate), that is with the disk formation scenario. At the
same time, the abundances of $\alpha$ and Fe-peak elements and their ratios to
Fe are crucial for getting 
insight into the role of stars with different masses and
evolutionary lifetimes in the heavy element enrichment of the interstellar
medium. 

The distribution of heavy elements with Galactocentric distance is often
investigated through observations of objects like H~{\sc ii} regions and B
stars (e.g. Shaver et al.~\cite{shaver};  Smartt \& Rollerstone \cite{smartt}),
which are bright and measurable also in external galaxies. They show a 
negative gradient, but it is not clear if the slope changes with  position in
the disk   and, given the young ages of all these objects, it is impossible to
evaluate whether it varies with age. Good indicators  are also planetary
nebul\ae~(PNe); observations of type~{\sc ii} PNe (whose progenitors are
$\sim$2--3 Gyr old) reveal the presence of a gradient similar to that traced by
H~{\sc ii} regions, but its precise slope is still under debate (e.g. Pasquali
\& Perinotto \cite{perinotto}; Maciel, Costa, \& Uchida \cite{maciel}),
since the investigation 
of PNe is affected by the uncertainty on their distances.
In our
opinion, one of the best tools for the investigation of radial  abundance
distributions is represented by open clusters, since
 they cover a wide range of ages,
metallicities, and positions in the Galactic disk.

Various observational studies have already addressed the problem of the metallicity
distribution through  open cluster observations, but the general picture 
has not been
delineated well yet, and discrepant results have
been obtained by  different authors.
For example, Friel (\cite{friel95}; \cite{friel_cast}),  Carraro, Ng, \&
Portinari (\cite{carraro98}) and Friel et al.~(\cite{friel02}) suggest the
presence of a negative [Fe/H] gradient, while Twarog, Ashman, \& Anthony-Twarog
(\cite{twarog}) and Corder \& Twarog
(\cite{corder})  favor a step-like  distribution of the Fe content with
Galactocentric distance. Furthermore, recent results (e.g., Yong
et al. \cite{yong05}) suggest that a single slope is not a good fit
to the data when clusters farther than about 14 kpc from the center are
included. Even when restricted to metallicities derived by spectroscopic data,
large differences appear to be present among different analyses.

The discrepancies among the various literature results can be ascribed to
several concurring factors: the number of stars employed (statistics), the
quality of the spectra (signal-to-noise -- $S/N$ -- and spectral resolution)
and, above all, the method of analysis, including continuum tracing,
atomic parameters, equivalent width ($EW$) measurement, model atmospheres,
spectral code, and
atmospheric parameters. Diverse methods of analysis  and
physical assumptions are adopted by different authors, so that the resulting
chemical abundances have been
 often  referred to discrepant scales of temperature and
metallicity. An investigation of this kind instead needs to rely on a
large sample of open clusters for which  element abundances (but also distances
and ages)  are derived with the same method in order to avoid spurious results
due to an inhomogeneous analysis.

Although the number of open clusters studied with high-resolution spectroscopy
is steadily increasing, we still lack a suitable sample for deriving truly
reliable constraints on the gradient and its evolution with time. To this aim
and
in the context of a VLT/FLAMES program on Galactic open clusters (Randich et
al.~\cite{messenger}), we collected UVES spectra of  evolved stars in a variety
of old open clusters. Our sample includes
\object{To~2} ($\sim$2--3 Gyr), 
\object{NGC~6253} ($\sim$3 Gyr), 
\object{Be~29} ($\sim$4 Gyr), \object{Be~20} ($\sim$5 Gyr),
\object{Be~32} ($\sim$6--7 Gyr)
and \object{NGC~2324}, \object{NGC~2477}, \object{NGC~2660},
\object{NGC~3960}
(ages $\lesssim$ 1 Gyr). Likewise $R_{\rm {gc}}$ varies between 
$\sim$7
and 21 kpc (Be~29 being the most distant cluster found so far).  The primary
goal of the UVES observations is to determinate the metallicity 
of the sample clusters  using a homogeneous method of analysis: in this
way, we will have a new rather large sample with all the clusters on the same
abundance scale. At the same time we derive -- in most cases for the first
time -- abundances of other key elements,  as CNO, $\alpha$, and Fe-peak
elements,  as well as s- and r-process elements, which are fundamental for 
understanding Galactic formation and evolution.

The emphasis of this paper is on presenting
the method of analysis for [Fe/H] and on determining 
the metallicity of three of the open clusters included in the
project (NGC~3960, NGC~2660, Be~32).
The analysis of $\alpha$ elements and Fe-peak
elements in the same clusters 
and in NGC~2477, NGC~2324
is deferred to a forthcoming paper (Bragaglia et
al.~2006b, in preparation). NGC~3960 was recently
investigated through high-resolution spectroscopy by Bragaglia  et
al.~(\cite{bragaglia06} -- hereafter B06a), therefore it is well-suited to
directly comparing it
with another spectroscopic analysis and to estimating possible
sources of discrepancies between different studies at high resolution.
As far as the other clusters are concerned, 
neither NGC~2660 nor Be~32
have ever been studied using high-resolution spectroscopy.

The paper is organized as follows: in Sect.~2 we present the properties of the
target clusters and in Sect.~3 we describe observations and data reduction. In
Sect.~4 we give a detailed description of the method of analysis and inputs
adopted (e.g., line lists and atomic parameters). The results for Fe abundances
in the three open clusters are reported  in Sect.~5 and discussed in Sect.~6. A
summary closes this paper (Sect.~7).

\section{Target clusters}

\begin{table*}[!] \footnotesize
\caption{Target clusters and their properties.
Literature sources are reported in the text.}\label{targets}
\begin{tabular}{llllllll}
\hline
\hline
Cluster & Age & [Fe/H] & $R_{\rm{gc}}$ & $(m-M)_{0}$ &$E(B-V)$ \\
        & (Gyr)&       & (kpc)  &(mag)   & (mag)  \\
\hline
NGC~3960 & 0.9$\rightarrow$1.4 &$-$0.68$\rightarrow$$-$0.06 &7.80 &11.60  & 0.29 (differential)\\
NGC~2660 & $\sim$1&solar/sub-solar &8.68 &12.20 &0.40 \\
Be~32 &$\sim$6$\rightarrow$7 &$-$0.50$\rightarrow$$-$0.37 &11.30 &12.48 &0.10 \\
\hline
\hline
\end{tabular}
\end{table*}

The main properties of the three open clusters investigated in this paper
are reported in Table~\ref{targets}.

NGC~3960 is located at a Galactocentric distance $R_{\rm{gc}}=7.8$ kpc. The
first CMD for the cluster was published by Janes (\cite{janes81}), based on
$BV$ photographic data. More recently, Prisinzano et al.~(\cite{prisinzano}) 
investigated this cluster by collecting  data in the $BVI$ bands with the Wide
Field Imager at the ESO-Max Planck 2.2m telescope. These authors found
strong indications of differential  reddening in the direction of the cluster,
with $E(B-V)$ varying from 0.16 to 0.62 over their $\approx$ 30 $\times$ 30
arcmin$^{2}$ field of view. They found  $E(B-V)=0.29$ towards the cluster
center (in agreement with Janes \cite{janes81}),  $(m-M)_{0}=11.35$, and age
ranging from 0.9 to 1.4 Gyr. The most recent work on NGC~3960 is the
photometric and spectroscopic study by B06a, who find $(m-M)_{0}=11.60$ and an
average $E(B-V)=0.29\pm0.02$ (over a field of view of 13.3 $\times$ 13.3
arcmin$^{2}$), with differential reddening of $\pm$0.05 mag (in agreement with
the values inferred by Prisinzano et al.~in their central region, corresponding
to B06a field of view) and a  slightly sub-solar metallicity, [Fe/H]=$-$0.12. 
This is the only metallicity estimate based on high-resolution spectroscopy.
Previous reports suggested a lower Fe content: for example, Friel \& Janes
(\cite{fj93})  quoted [Fe/H]=$-$0.34
from low-resolution spectroscopy. Other 
estimates range from [Fe/H]=$-$0.68 (Geisler, Clari\`a, \& Minniti
\cite{geisler}; from Washington photometry) to $-$0.06 (Piatti, Clari\`a, \&
Abbadi \cite{piatti},
using a DDO abundance calibration). 
NGC~3960 has been included in several other studies; for
example, by Twarog et al.~(\cite{twarog}),  who determined a metallicity of
$-$0.17, based on a homogenization of DDO photometry and low-resolution
spectroscopy, and by Mermilliod et al.~(\cite{mermi01}), who determined precise
radial velocities for a number of red giants.

NGC~2660 was first investigated by Hartwick \& Hesser (\cite{hh73};
photographic and photoelectric study), and a more recent analysis was performed
by Sandrelli et al.~(\cite{sandrelli}), who presented CCD $UBVI$ photometry
and estimated an age of $\sim$ 1 Gyr or slightly less, a distance modulus
$(m-M)_{0}=12.20$, and reddening $E(B-V)\sim0.40$, implying a  Galactocentric
distance of 8.68 kpc (Bragaglia \& Tosi \cite{bt06}). The cluster metallicity
has been investigated by various authors but only on the basis of photometry,
and with inconsistent results: Sandrelli et al.~(\cite{sandrelli}) quoted a
nearly solar Fe content (from stellar evolutionary
tracks), while previous works reported a sub-solar [Fe/H]
(e.g., Hesser \& Smith \cite{hs87}; Piatti et al ~\cite{piatti},
both from DDO photometry).

As pointed out in Sect.~1, the main
goal of this project is to investigate the radial metallicity
gradient in the disk and its temporal evolution. 
Be~32 is $\sim$6--7 Gyr old, which is relevant for the evolution
of the gradient with age. Although the most distant cluster
in our sample is Be~29 ($R_{\rm{gc}}\sim21$ kpc), Be~32
is the most distant  among the three clusters analyzed in this
paper and, with a Galactocentric radius $>$ 10 kpc,
it is located beyond the [Fe/H] vs. $R_{\rm{gc}}$ discontinuity 
found by Twarog et al.~(\cite{twarog}).
Therefore,
Be~32 is the most interesting of the three clusters. 
Photometric studies of the cluster were carried out by Kaluzny \&
Mazur (\cite{kal}), Richtler \& Sagar (\cite{rich}), Hasegawa et
al.~(\cite{hasegawa}) and D'Orazi et al.~(\cite{dorazi}). The last study
finds reddening and distance modulus $E(B-V)=0.10$ and $(m-M)_{0}=12.48$.  As
for the metal content, a sub-solar Fe has been suggested for the cluster, e.g.,
by Noriega-Mendoza \& Ruelas-Mayorga (\cite{noriega}), who found [Fe/H]=$-$0.37
based on the CMD, by Friel et al.~(\cite{friel02})  based on low-resolution
spectroscopy ([Fe/H]=$-$0.50), and by D'Orazi et al.~(\cite{dorazi}) based on
evolutionary tracks ($Z$=0.008). 

\section{Observations and data reduction}\label{obs}

The open clusters included in the program were all observed  with the
multi-object instrument FLAMES on VLT/UT2 (ESO, Chile; Pasquini et
al.~\cite{P00}). The fiber link to UVES was used to obtain high-resolution
spectra ($R=40,000$) for red giant branch (RGB) and clump objects.

NGC~3960 was observed in Service mode in 2004; two FLAMES configurations were
used, and UVES observations were  performed for both configurations with two
different gratings (CD3 and CD4, covering the wavelength ranges
$\sim$4750--6800 \AA~and 6600--10600 \AA, respectively). A log of observations
(date, UT, exposure time, grating, configuration, number of stars)  is given in
Table~\ref{obs3960}. The spectra were reduced by ESO personnel using the
dedicated pipeline, and we analyzed the 1-d, wavelength-calibrated spectra
using standard  IRAF\footnote{IRAF is distributed by the National Optical
Astronomical Observatories, which  are operated by the Association of
Universities for Research in Astronomy,  under contract with the National
Science Foundation.}  packages. 

Table~\ref{data3960} provides information on the target stars. We observed  7
red clump and 3 main sequence (MS) stars, but we only present here the results
for the giants. We adopt hereafter  the provisional identification number used
for FLAMES pointing (Col.~1); however, since $BVI$ photometry  was taken from
Prisinzano et al.~(\cite{prisinzano}), we also
show in Col.~2  the ID from their
catalogue. $J$ and $K$ magnitudes come from the Two Micron All Sky Survey
(2MASS\footnote{The Two Micron All Sky Survey is a joint project of the
University of Massachusetts and the Infrared Processing and Analysis
Center/California Institute of Technology, funded by the national Aeronautics
and Space Administration and the National Science Foundation.}; Cutri et
al.~\cite{cutri}). The MS stars were eventually discarded
since they are rather faint, 
too warm (late A or early F spectral type), 
and rotate too rapidly. They
are also not suitable
for a detailed chemical abundance analysis,
since the $S/N$ is too low
and their lines are broad and shallow.

Radial velocities were measured using RVIDLINES on several tens of metallic
lines on the individual spectra, then multiple spectra were combined. One
fiber for configuration was used to register the sky value, but the correction
was  negligible. We corrected the spectra for  the contamination by atmospheric
telluric lines using TELLURIC in IRAF and an early-type star observed with UVES
for another program.  The $RV$s for each star (shown in Table 3) have an
attached uncertainty of less than about 0.5 km~s$^{-1}$, as deduced from the
rms when averaging values obtained from different exposures; this error is
also valid for the two other clusters.
We derived an average heliocentric $RV$ of $-22.6\pm0.9$  (statistical error)
km~s$^{-1}$ for NGC~3960. Star c1 turned out to be a non-member on the basis of
the radial velocity and spectral characteristics and is disregarded from now on
(see also Mermilliod et al.~\cite{mermi01}). Star c8 is a long-period
spectroscopic binary (star 50
in Mermilliod et al.~\cite{mermi01}), so
its radial velocity appears slightly discrepant
with respect to the cluster average; nevertheless it is compatible with it,
once the amplitude of the $RV$ curve is considered. The spectral features and
average metallicity (see Sect.~\ref{risultati}) clearly indicate that
c8 is a
cluster member.

\begin{table*}[!] \footnotesize
\caption{Observation log of NGC~3960.
One star is common to configurations A and B;
we consider only clump objects here (7 stars in total). The remaining
fibers were assigned to sky and to MS stars.}\label{obs3960}
\begin{tabular}{llllll}
\hline
\hline
Date   & UT$_{\rm{beginning}}$ &Exposure time&Configuration &Grating& no. of stars\\
& &(s) & && (clump)\\
\hline
2004-04-03  &  02 57 39.980   & 2595& A      & CD4 & 4\\
2004-04-03  &  03 52 16.186   & 2595&A      & CD4 &4\\
2004-04-03  &  04 47 31.818   & 2595&A      & CD3 &4\\
2004-04-20  &  03 37 39.635   & 2595&B      & CD3 &4\\
2004-05-02  &  00 12 31.740   & 2595&B      & CD4 &4\\
2004-05-02  &  00 57 37.007   & 2595&B      & CD4 &4\\
2004-05-02  &  01 43 13.805   & 2595&B      & CD3 &4\\
2004-05-02  &  02 39 18.200   & 2595&A      & CD3 &4\\
\hline
\hline
\end{tabular}
\end{table*}

\begin{table*}[!] \footnotesize
\caption{Data for NGC 3960. ID$_{\rm{phot}}$ and
$BVI$ photometry (non corrected for differential reddening)
are from Prisinzano et al.~(\cite{prisinzano}).The number of exposures
for each star is intended as the number of pointings
with the same cross-disperser. }\label{data3960}
\begin{tabular}{llllllllllllll}
\hline
\hline
Star & Star  &   RA	     &     DEC      & $B$      &  $V$     &   $I$    & $J_{\rm{2MASS}}$  & $K_{\rm{2MASS}}$ & no. exp. & $RV$   & $S/N$&Notes \\
ID$_{\rm{fla}}$&ID$_{\rm{phot}}$&&&&&&&&&(km s$^{-1}$)& &\\
\hline
   c1 &310753 & 11 50 06.330 & $-$55 44 13.00 & 15.866 & 14.262 & 12.496 & 11.213  &10.238  &2 & +1.83 & 60--80 &NM  \\
   c3 & 310755 & 11 50 33.149 & $-$55 42 36.13 & 14.799 & 13.514 & 12.088 & 11.079  &10.357  &2 &$-$24.16  & 120--130& M	 \\
   c4 & 310756 & 11 50 28.184 & $-$55 41 36.66 & 14.366 & 13.194 & 11.907 & 10.989  &10.304  &2 &$-$22.39  & 95--115& M	 \\
   c5 & 310757 & 11 50 36.050 & $-$55 42 05.40 & 14.323 & 13.062 & 11.679 & 10.671  & 9.959  &2 &$-$21.94 & 110--130& M	 \\
   c6 & 310758 & 11 50 26.750 & $-$55 40 28.20 & 14.084 & 12.945 & 11.697 & 10.815  &10.133  &4 &$-$21.86  & 115--160& M	 \\
   c8 & 310760 & 11 50 37.621 & $-$55 40 15.19 & 14.224 & 13.060 & 11.758 & 10.820  &10.162  &2 &$-$32.87  & 120--190 & bin, M\\
   c9 & 310761 & 11 50 38.100 & $-$55 39 44.50 & 14.309 & 13.100 & 11.787 & 10.839  &10.151  &2 &$-$22.58  & 100--110& M	 \\
\hline
\hline
\end{tabular}
\end{table*}

Be 32 and NGC 2660 were observed in Visitor mode in 2005 January 19--22,
and a log
of the observations is given in Table~\ref{obs32_2660}.  Be~32 was targeted
using two FLAMES configurations,
For configuration A, the observations were
carried out with both CD3 and CD4 gratings, while only CD3
was employed for configuration B  For NGC~2660, 
we got only one exposure with the CD3 grating.
The spectra were reduced by us, using the dedicated UVES pipeline (Mulas et
al.~\cite{uves_pipeline}). Since observations were performed at full   Moon and
in the 
presence of cirrus, the background signal is high; nevertheless, we were able
to carry out a good background subtraction using, as is
customary, the fibers
dedicated  to the sky.  Information on radial velocities and photometry are
recorded in Tables~\ref{data32} and \ref{data2660} for Be~32 (10 stars) and
NGC~2660 (5 stars), respectively.

In Table~\ref{data32} we report the provisional ID for the stars used for the
FLAMES pointing  (adopted in the following) and those from the photometry by 
D'Orazi et al.~(\cite{dorazi}). Note that for star 938  the photometry was
retrieved from Kaluzny \& Mazur (\cite{kal}) and that star 932 turned out to
be a non-member, because of both its $RV$ and its spectral
features; therefore it  has been dropped from the analysis.
Identification numbers and $BVI$ magnitudes for NGC~2660 (listed in
Table~\ref{data2660})  were taken from Sandrelli et al.~(\cite{sandrelli}).

The average heliocentric $RV$s for Be~32 and NGC~2660 are $<RV>=+106.0\pm1.4$
km~s$^{-1}$ and $<RV>=+21.2\pm0.6$ km~s$^{-1}$, respectively. The $S/N$ given
in Tables 3, 5, 6 refers to the spectra centered at 5800 \AA, on which our
analysis rests, measured using the task SPLOT within IRAF. Considering the
wavelength regions around $\sim$5600 and 6300 \AA, the final combined spectra
of NGC~3960 have typical $S/N\sim$100--160; lower values result for Be~32,
which has $S/N\sim$50--100. Finally, spectra of NGC~2660, for which only one
exposure for each star was available, have typical $S/N$ of $\sim$45--70.

\begin{figure*}
\includegraphics[bb=0 400 592 718, clip, scale=0.8]{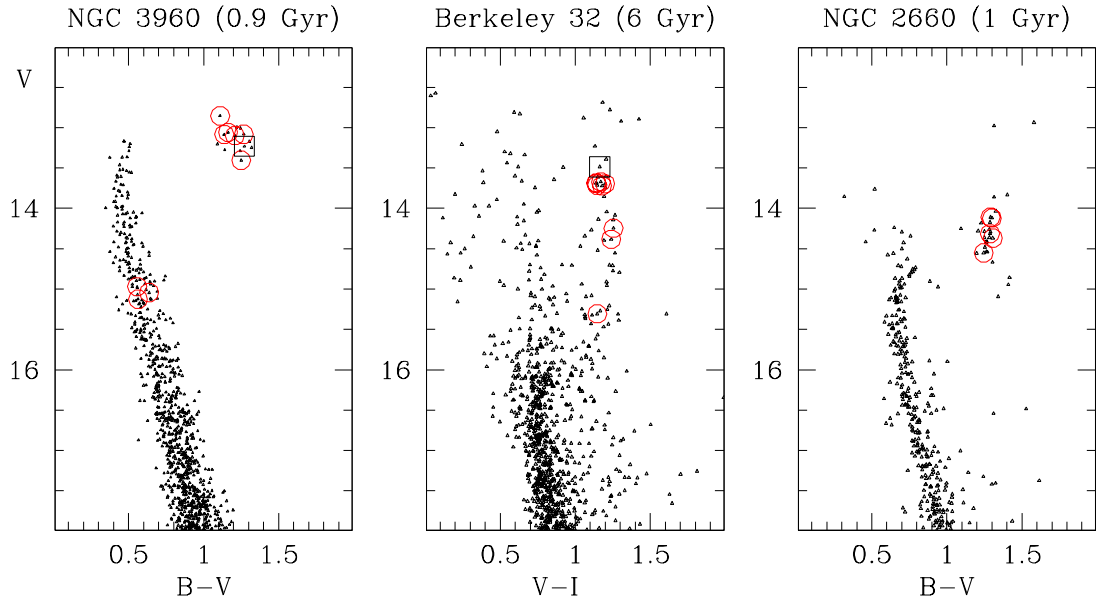}
\caption{Color-magnitude diagrams for the three clusters. From left to right:
NGC~3960, Be~32, and NGC~2660. Note that
the first and the last clusters are reported in the ($B-V$, $V$) plane, while
Be~32 is in the ($V-I$, $V$) one. The observed stars are shown
by circles (members) and squares (non-members).}\label{CMD}
\end{figure*}

Figure~\ref{CMD} shows the CMDs for the three open clusters in the ($B-V$,
$V$) plane for NGC~3960 and NGC~2660, while for Be~32 we considered the ($V-I$)
colors because the $B$ magnitude of star 938 is not available. Circles
(members) and squares (non-members) mark the stars observed with UVES.

\begin{figure*}
\psfig{figure=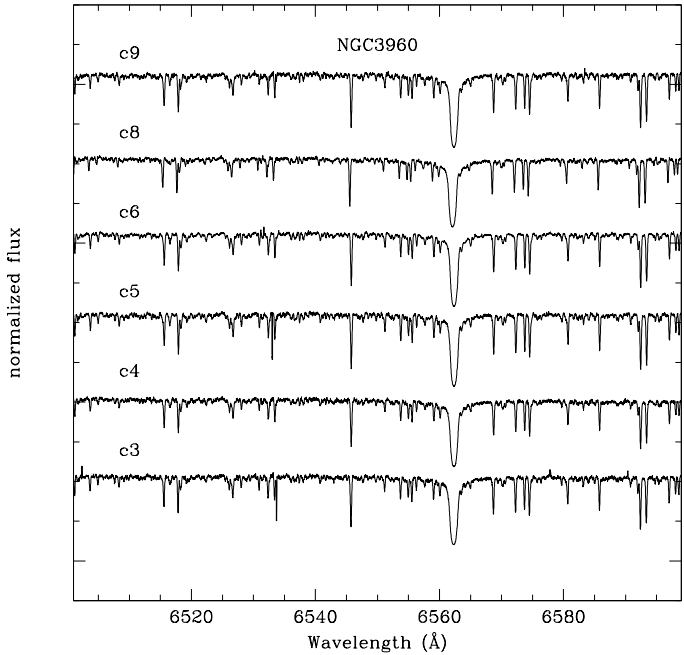, width=10cm, angle=0}
\caption{NGC~3960 sample spectra in the spectral region
at 6500--6600 \AA.}\label{sp3960}
\end{figure*}

\begin{figure*}
\psfig{figure=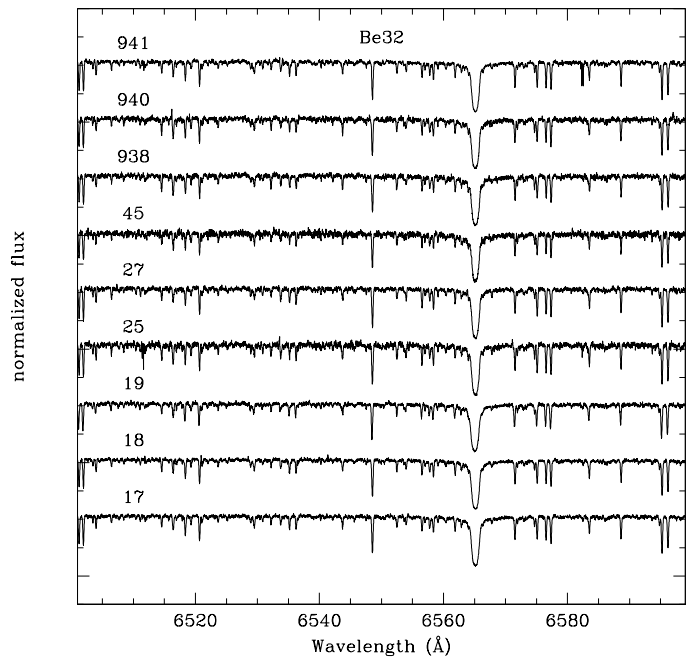, width=10cm, angle=0}
\caption{Be~32 sample spectra.}\label{sp32}
\end{figure*}

\begin{figure*}
\psfig{figure=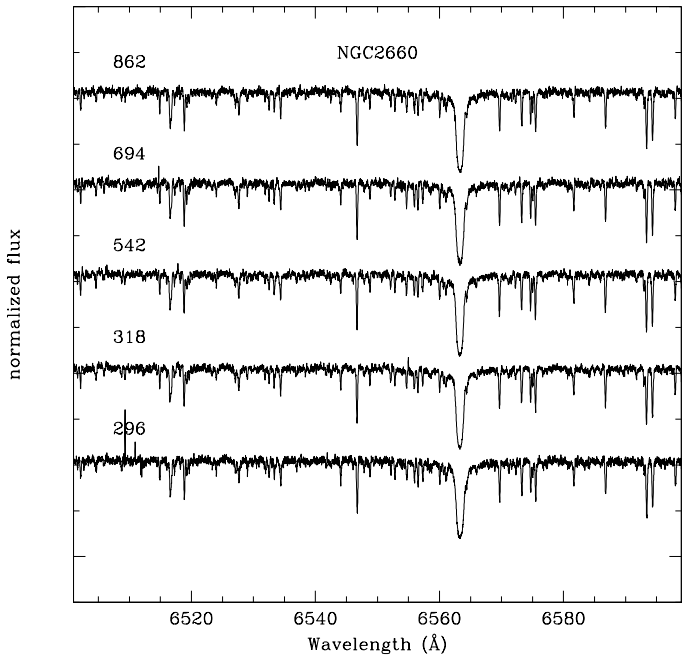, width=10cm, angle=0}
\caption{NGC~2660 sample spectra.}\label{sp2660}
\end{figure*}

We show  the spectra of all
the clump stars in the sample (cluster members)
in Figs.~\ref{sp3960}, \ref{sp32},
 and \ref{sp2660}. For clarity we restricted the
plot to a region of $\sim$100 \AA~at 6500--6600 \AA.

\begin{table*}[!] \footnotesize
\caption{Observation log of Be~32 and NGC~2660. One star is common to
configurations A and B in Be~32 (10 stars in total). The remaining fibers were
assigned to the sky.}\label{obs32_2660}
\begin{tabular}{llllllll}
\hline
\hline
 Cluster  &    Date    &  UT$_{\rm{beginning}}$  & Exp.time & Config.&  Grating&no. of stars \\
&&&(s)&&& \\
\hline
Be 32     & 2005-01-20 & 00 47 00 &  3600 & A & CD3 & 7 \\
Be 32     & 2005-01-20 & 02 01 53 &  3600 & A & CD4 & 7  \\
Be 32     & 2005-01-20 & 03 11 03 &  3600 & A & CD4 & 7 \\
Be 32     & 2005-01-20 & 04 23 41 &  3600 & A & CD3 & 7 \\ 
Be 32     & 2005-01-21 & 00 45 26 &  3600 & B & CD3 & 4 \\ 
NGC 2660  & 2005-01-23 & 08 02 02 &  3600 & --& CD3 & 5 \\
\hline
\hline
\end{tabular}
\end{table*}

\begin{table*}[!] \footnotesize
\caption{Data for Be 32. ID$_{\rm{phot}}$, and $BVI$ are from D'Orazi et
al.~(\cite{dorazi}), while ID$_{\rm{fla}}$ is the identification number used
for FLAMES observations.}\label{data32}
\scriptsize
\begin{tabular}{lllllllllllllll}
\hline
\hline
\scriptsize
Star & Star &      RA    &     DEC    &   $B$ &   $V$ &	$I$ &  $J_{\rm{2MASS}}$  & $K_{\rm{2MASS}}$ & no. exp.& $RV$ & $S/N$ & Notes\\
ID$_{\rm{fla}}$& ID$_{\rm{phot}}$&&&&&&&&&(km s$^{-1}$)&&\\
\hline
    17 &  533  & 6 58  8.242 & 6 24 19.48 & 14.754 & 13.667 & 12.540 & 11.683 & 11.028 &2& 105.3 &80--100&M\\
    18 &  997  & 6 58 13.762 & 6 27 54.89 & 14.730 & 13.661 & 12.534 & 11.709 & 11.069 &2& 105.5 &80--100&M\\
    19 &  787  & 6 58  3.089 & 6 26 16.08 & 14.800 & 13.709 & 12.564 & 11.722 & 11.024 &2& 101.4 &70--90 &M\\
    25 &  121  & 6 57 59.819 & 6 26 59.93 & 15.387 & 14.242 & 13.028 & 12.153 & 11.450 &1& 105.5 &45--60 &M\\
    27 &  605  & 6 58  2.262 & 6 24 56.71 & 15.511 & 14.372 & 13.177 & 12.282 & 11.607 &2& 105.6 &60--70 &M\\
    45 & 1139  & 6 58  7.473 & 6 29 32.67 & 16.299 & 15.278 & 14.153 & 13.341 & 12.707 &3& 105.3 &45--50 &M\\
   932 & 1183  & 6 58  1.771 & 6 29 55.32 & 14.450 & 13.425 & 12.282 & 11.474 & 10.843 &2&  30.8 &90--100&NM\\
   938 &   -   & 6 58 11.488 & 6 21 16.25 & -      & 13.672*& 12.501*& 11.686 & 11.027 &1& 106.1 &50--60 &M\\
   940 &  104  & 6 58 22.909 & 6 26 25.26 & 14.793 & 13.691 & 12.535 & 11.691 & 11.023 &1& 105.0 &70--75 &M\\
   941 &   99  & 6 57 50.572 & 6 26 11.92 & 14.769 & 13.663 & 12.477 & 11.643 & 10.973 &2& 105.5 &85--95 &M\\
\hline
\hline
\end{tabular}

*$V$ and $I$ from Kaluzny \& Mazur (\cite{kal}).
\end{table*}

\begin{table*}[!] \footnotesize
\caption{Data for NGC 2660. ID and $BVI$ photometry 
are from Sandrelli et al.~(\cite{sandrelli}).}.\label{data2660}
\begin{tabular}{lllllllllllll}
\hline
\hline
Star   & RA          &  DEC        & $B$      &  $V$     &  $I$     & $J_{\rm{2MASS}}$      &  $K_{\rm{2MASS}}$     &  $RV$  & $S/N$& Notes\\
&&&&&&&&(km s$^{-1}$) & &\\
\hline
 296  & 8 42 36.411 & $-$47 12 07.26 & 15.798 & 14.552 & 13.174 & 11.680 & 10.636 &  21.73 & 45--65&M\\ 
 318  & 8 42 36.693 & $-$47 10 37.06 & 15.401 & 14.110 & 12.713 & 11.586 & 10.828 &  20.66 &50--65&M\\ 
 542  & 8 42 41.704 & $-$47 11 25.22 & 15.420 & 14.120 & 12.704 & 11.603 & 10.837 &  20.47 &45--90&M\\
 694  & 8 42 45.418 & $-$47 11 18.71 & 15.674 & 14.368 & 12.938 & 11.791 & 11.049 &  21.47 &45--55&M\\  
 862  & 8 42 50.651 & $-$47 13 03.61 & 15.602 & 14.315 & 12.909 & 11.770 & 11.046 &  21.47 & 50--70&M\\  
\hline
\hline
\end{tabular}
\end{table*}

\section{Abundance analysis}\label{analysis}

The analysis of chemical abundances was carried out with the  last version
(2005) of the spectral program MOOG (Sneden
\cite{sneden})\footnote{http://verdi.as.utexas.edu/} and using model
atmospheres by Kurucz (\cite{kuru}).  Like all the commonly
used spectral analysis codes, 
MOOG performs a local thermodynamic equilibrium
(LTE) analysis.

\subsection{Solar analysis}\label{sun}

The first step is the determination of the solar Fe abundance,  which allows us
to fix a zero-point for the metallicity scale. This differential analysis
minimizes errors in the results, especially for stars with nearly solar
metallicity.

The line list adopted for the Sun is the one used
by Gratton et al.~(\cite{G03} -- hereafter G03), which includes 180 and 40
features for Fe~{\sc i} and Fe~{\sc ii}, respectively, in the wavelength range
$\sim$4100--6800 \AA. Either theoretical or laboratory oscillator strengths
($\log gf$) were considered, retrieved from the works by the Oxford group (e.g.
Blackwell et al.~\cite{oxford}), and from Bard, Kock, \& Kock (\cite{bard91}),
O'Brian et al.~(\cite{obrian91}), 
and Bard \& Kock (\cite{bard94}). Most of the
equivalent widths adopted for the Sun  are from Rutten \& Van der Zalm
(\cite{rutten}), who performed measurements on the Sacramento Peak Irradiance
Atlas (Beckers, Bridges, \& Gilliam \cite{sacramento}); this set was integrated
with measurements performed by  G03 on the solar spectrum atlases by
Delbouille, Roland, \& Neven (\cite{atlas_del}) and by
Kurucz~(\cite{atlas_kuru}). Note that the solar spectra employed have much
higher resolution than our UVES spectra. 

When available, we adopted collisional damping  coefficients from Barklem,
Piskunov, \& O'Mara (\cite{barklem}),  who provide  the best theoretical models
for most transitions\footnote{MOOG was recently updated in order to give the
possibility of using the Barklem coefficients.}. When the coefficients by
Barklem et al.~were not available we considered classical damping constants
($C_{6}$) computed with the approximation of Uns\"old (\cite{unsold})  and
multiplied  by an enhancement factor E, given by:  $\log
E=(0.381\pm0.017)EP-(0.88\pm0.33)$. This expression was obtained by G03 from
several hundred Fe~{\sc i} features with available accurate collisional
damping parameters and for \teff=5000 K, typical of giant stars. 

The line list for the Sun is available in electronic form;
the table includes wavelengths, name of the element (Fe~{\sc i}
or Fe~{\sc ii}), $EP$, $\log gf$, $EW$, and 
references for the damping coefficients.
We computed the Fe content ($\log n(\rm{Fe})$\footnote{$\log n(\rm{Fe})=12+\log
(N(\rm{Fe})/N(\rm{H}))$, absolute number density abundance.})  for the Sun
by adopting the following effective temperature, surface gravity, and
microturbulence velocity: \teff=5779 K, $\log g$=4.44, and $\xi$=0.8 km
s$^{-1}$. The values of the solar microturbulence reported in the literature
range from 0.8 to 1.1 km s$^{-1}$ (e.g. Randich et al.~\cite{R06M67}).
We chose
0.8 km s$^{-1}$, which is the best fit value reported by Grevesse \& Sauval
(\cite{GS99}); as well known, an increase in $\xi$ would result into a decrease
of  the solar $\log n(\rm{Fe})$. We derived $\log n(\rm{Fe}$~{\sc
i})=7.49$\pm$0.04 (standard deviation, or rms)  and $\log n(\rm{Fe}$~{\sc
ii})=7.54$\pm$0.03, in good agreement with the estimates available in the
literature (e.g., Grevesse \& Sauval \cite{GS99}; Asplund, Grevesse, \& Sauval
\cite{AS05}). Usual checks for the suitability of the adopted parameters are
the excitation equilibrium and the plot of Fe abundance vs.~$EW$s (see
Sect~\ref{par}). We find a negligible trend of the Fe~{\sc i} abundance with
the excitation potential ($EP$) -- hence the excitation equilibrium is
satisfied (see Sect.~\ref{par}) -- while a slightly stronger trend is present
considering $\log n(\rm{Fe}$~{\sc i}) vs.~the measured $EW$s (a positive slope
of $\sim$0.06 in the plot of abundance vs.~$\log (EW/{\lambda})$). In order to
put the value of this slope to $\sim$0, we should have slightly enhanced the
microturbulence (up to 0.95 km s$^{-1}$) and, as a consequence, the \teff~(up to
5800 K); in this case, the Fe abundance would have been 7.48.
However, we decided to retain the initial parameters
for the Sun as the right ones and therefore did not put
the slopes of $\log n(\rm{Fe}$~{\sc i}) vs.~measured $EW$s
and vs.~$EP$ to zero.
We warn the reader that this choice  can lead to a systematic shift in the
metallicity scale; had we assumed a lower $\log n(\rm{Fe}$~{\sc i}), we would
have found slightly higher metallicities for the open cluster stars.

Finally, we note that using the previous version (2002) of MOOG, where it is
not possible to treat collisional damping with the Barklem coefficients, we
would have obtained $\log n(\rm{Fe}$~{\sc i})=7.51$\pm$0.04 and $\log
n(\rm{Fe}$~{\sc ii})=7.55$\pm$0.04 (considering classical damping constants
from the Uns\"old formula, in most cases multiplied by the enhancement factor
$E$).

\subsection{Line list for giant stars}\label{giants}

The line list adopted for giant stars in open clusters (retrieved from G03)
covers the wavelength range 5500--6800 \AA~for Fe~{\sc i} (157 lines) and
5500--6500 \AA~for Fe~{\sc ii} (15 lines). The Fe features included were first
carefully selected by G03 in order to exclude the presence of severe blending.
Moreover, very strong lines ($EW\gesssim150$ m\AA, see Sect.~\ref{ews})
have also been discarded, since they are critically
sensitive to the microturbulence value; and further, a more detailed treatment
of damping  would be needed to fit the line wings. Note that, although the
investigation by G03 concerned metal-poor halo stars, the line list adopted
here was accurately tested and employed for solar metallicity open clusters
(Carretta et al.~\cite{carretta04}, \cite{carretta05}; B06a).
Fe features at wavelengths bluer than $\sim$5500 \AA~were excluded
to avoid complications by crowding and continuum tracing:
therefore, $\sim$60 Fe~{\sc i}  and $\sim$25
Fe~{\sc ii} lines
 used for the solar analysis are not included
in the giant list. On
the other hand,  there are additional useful
Fe~{\sc i} features with respect to the solar line list,
in the  list used for giants (since they are visible only
in cool stars). 
Based on these facts, the analysis is not
formally ``differential'' with respect to the Sun; however,  as a test we
measured the solar abundance using the lines shared by the two lists
and found the same $\log n(\rm{Fe})$ values, reinforcing our method.
More in detail,
the fraction of Fe~{\sc i} features
not included in the solar line list are about 25 \% of the lines used for
giants. This, in principle, might
represent a systematic bias, since
the excitation equilibrium of Fe~{\sc i}   lines
(and therefore the \teff) 
is strongly driven by the lowest and highest excitation
lines.
However, 
the majority of the features not shared by the Sun and giants
have $EP$ in the range 3--5 eV,
similar to those of the other lines present in the total lists (see
also below). Only two
lines 
in this  subsample have very low $EP$ ($<2$ eV);
therefore, the use of two different subsets of lines for the Sun
and the giants does not represent a source of error.

Wavelengths, $EP$, and $\log gf$ for the lines
used are reported in Table~\ref{linelist}. A wide
range in $EP$ and $\log gf$ (i.e., corresponding to a wide range of line
strengths) is spanned at all wavelengths by Fe~{\sc i} features, with more than
70 \% of the lines having $EP\geq3$ eV and ${\log gf\geq-3}$; only 10 features
have $EP$ lower than 2 eV and $\log gf$ lower than $-$4.

The sources for the treatment of damping are also reported: G03 (see
Sect.~\ref{sun}), Barklem et al.~(\cite{barklem}), Uns\"old (\cite{unsold}) .
Note that only in two cases (where the Barklem and G03 values were not
available) was the classical Uns\"old approximation used.

\setcounter{table}{6}
\begin{table*}[!] \footnotesize
\caption{Set of Fe lines adopted for the analysis.}\label{linelist}
\scriptsize
\begin{tabular}{lllllllllll}
\hline
\hline
Wavelength &$EP$ &$\log gf$ &Damping& Wavelength &$EP$ &$\log gf$ &Damping\\
(\AA) & (eV) & & source& (\AA) & (eV) & & source\\
\hline
Fe I      &&&&                           5806.732 & 4.610 & -0.930 &	2  \\
 5494.474 & 4.070 & -1.960 &      1  &   5811.912 & 4.140 & -2.360 &	1 \\
 5521.281 & 4.430 & -2.510 &      1  &	  5814.815 & 4.280 & -1.810 &	2 \\
 5522.454 & 4.210 & -1.470 &   2  &	  5835.109 & 4.260 & -2.180 &	2 \\
 5524.244 & 4.150 & -2.840 &   2  &	  5837.702 & 4.290 & -2.300 &	2 \\
 5539.291 & 3.640 & -2.590 &   2  &	  5849.687 & 3.690 & -2.950 &	2 \\
 5547.000 & 4.220 & -1.850 &   2  &	  5852.228 & 4.550 & -1.360 &	2 \\
 5552.687 & 4.950 & -1.780 &   2  &	  5853.150 & 1.480 & -5.090 &	1 \\
 5560.220 & 4.430 & -1.100 &   2  &	  5855.086 & 4.610 & -1.560 &	2 \\
 5568.862 & 3.630 & -2.910 &   2  &	  5856.096 & 4.290 & -1.570 &	2 \\
 5577.028 & 5.030 & -1.490 &      1  &	  5858.785 & 4.220 & -2.190 &	2 \\
 5586.771 & 3.370 & -0.100 &   2  &	  5859.596 & 4.550 & -0.630 &	1 \\
 5587.581 & 4.140 & -1.700 &      1  &	  5861.110 & 4.280 & -2.350 &	2 \\
 5595.051 & 5.060 & -1.780 &      1  &	  5862.368 & 4.550 & -0.420 &	1 \\
 5608.976 & 4.210 & -2.310 &   2  &	  5879.490 & 4.610 & -1.990 &	2 \\
 5609.965 & 3.640 & -3.180 &   2  &	  5880.025 & 4.560 & -1.940 &	2 \\
 5611.357 & 3.630 & -2.930 &   2  &	  5881.279 & 4.610 & -1.760 &	2 \\
 5618.642 & 4.210 & -1.340 &   2  &	  5902.476 & 4.590 & -1.860 &	2 \\
 5619.609 & 4.390 & -1.490 &   2  &	  5905.680 & 4.650 & -0.780 &	2 \\
 5635.831 & 4.260 & -1.590 &   2  &	  5927.797 & 4.650 & -1.070 &	2 \\
 5636.705 & 3.640 & -2.530 &   2  &	  5929.682 & 4.550 & -1.160 &	2 \\
 5649.996 & 5.100 & -0.800 &   2  &	  5930.191 & 4.650 & -0.340 &	2 \\
 5651.477 & 4.470 & -1.790 &   2  &	  5933.805 & 4.640 & -2.140 &	2 \\
 5652.327 & 4.260 & -1.770 &   2  &	  5934.665 & 3.930 & -1.080 &	2 \\
 5661.017 & 4.580 & -2.420 &   2  &	  5947.531 & 4.610 & -2.040 &	1 \\
 5661.354 & 4.280 & -1.830 &   2  &	  5956.706 & 0.860 & -4.560 &	2 \\
 5677.689 & 4.100 & -2.640 &   2  &	  5976.787 & 3.940 & -1.300 &	1 \\
 5678.388 & 3.880 & -2.970 &      1  &	  5984.826 & 4.730 & -0.290 &	1 \\
 5680.244 & 4.190 & -2.290 &      1  &	  6003.022 & 3.880 & -1.020 &	2 \\
 5701.557 & 2.560 & -2.160 &   2  &	  6007.968 & 4.650 & -0.760 &	1 \\
 5717.841 & 4.280 & -0.980 &   2  &	  6008.566 & 3.880 & -0.920 &	1 \\
 5731.772 & 4.260 & -1.100 &   2  &	  6015.242 & 2.220 & -4.660 &	2 \\
 5738.240 & 4.220 & -2.240 &   2  &	  6019.369 & 3.570 & -3.230 &	2 \\
 5741.856 & 4.260 & -1.690 &   2  &	  6027.059 & 4.070 & -1.200 &	1 \\
 5742.963 & 4.180 & -2.350 &   2  &	  6056.013 & 4.730 & -0.460 &	2 \\
 5752.042 & 4.550 & -0.920 &      1  &	  6065.494 & 2.610 & -1.490 &	2 \\
 5754.406 & 3.640 & -2.850 &   2  &	  6078.499 & 4.790 & -0.380 &	1 \\
 5759.259 & 4.650 & -2.070 &   2  &	  6079.016 & 4.650 & -0.970 &	2 \\
 5760.359 & 3.640 & -2.460 &   2  & 	  6082.718 & 2.220 & -3.530 &	2 \\
 5775.088 & 4.220 & -1.110 &      1  &	  6089.574 & 5.020 & -0.870 &	1 \\
 5778.463 & 2.590 & -3.440 &   2  &	  6093.649 & 4.610 & -1.320 &	2 \\
 5784.666 & 3.400 & -2.530 &   2  &	  6094.377 & 4.650 & -1.560 &	2 \\
 5793.922 & 4.220 & -1.620 &   2  &	  6096.671 & 3.980 & -1.760 &	2 \\
\hline
\hline
\end{tabular}

References: (1) Gratton et al.~(\cite{G03}; G03);
(2) Barklem et al.~(\cite{barklem}); (3) Uns\"old (\cite{unsold}).
\end{table*}

\setcounter{table}{6}
\begin{table*}[!] \footnotesize
\caption{(continued).}\label{linelist2}
\scriptsize
\begin{tabular}{lllllllllll}
\hline
\hline
Wavelength &$EP$ &$\log gf$ &Damping& Wavelength &$EP$ &$\log gf$ &Damping\\
(\AA) & (eV) & & source& (\AA) & (eV) & & source\\
\hline
6098.250 & 4.560 & -1.810 &	2 &    6581.218 & 1.480 & -4.680 &   2 \\
6120.258 & 0.910 & -5.860 &	 1 &   6591.314 & 4.590 & -2.040 &   2 \\     
6137.002 & 2.200 & -2.910 &   2 &      6593.884 & 2.430 & -2.300 &   2 \\     
6151.623 & 2.180 & -3.260 &   2 &      6608.044 & 2.280 & -3.960 &   2 \\     
6157.733 & 4.070 & -1.260 &	 1 &   6609.118 & 2.560 & -2.650 &   2 \\     
6165.363 & 4.140 & -1.480 &	 1 &   6625.039 & 1.010 & -5.320 &   1 \\   
6173.341 & 2.220 & -2.840 &   2 &      6627.560 & 4.550 & -1.500 &   2 \\     
6187.402 & 2.830 & -4.130 &	 1 &   6633.758 & 4.560 & -0.810 &   2 \\     
6187.995 & 3.940 & -1.600 &   2 &      6667.426 & 2.450 & -4.370 &   2 \\     
6199.509 & 2.560 & -4.350 &   2 &      6667.723 & 4.580 & -2.100 &   2 \\     
6200.321 & 2.610 & -2.390 &   2 &      6699.142 & 4.590 & -2.110 &   2 \\     
6213.437 & 2.220 & -2.540 &   2 &      6703.576 & 2.760 & -3.000 &   2 \\     
6219.287 & 2.200 & -2.390 &   2 &      6704.485 & 4.220 & -2.640 &   1\\   
6220.791 & 3.880 & -2.360 &   2 &      6713.745 & 4.790 & -1.410 &   2 \\     
6226.740 & 3.880 & -2.080 &   2 &      6725.364 & 4.100 & -2.210 &   2 \\     
6232.648 & 3.650 & -1.210 &   3&       6726.673 & 4.610 & -1.050 &   1\\   
6240.653 & 2.220 & -3.230 &   2 &      6733.153 & 4.640 & -1.440 &   2 \\     
6246.327 & 3.600 & -0.730 &   2 &      6739.524 & 1.560 & -4.850 &   2 \\     
6252.565 & 2.400 & -1.640 &   2 &      6745.965 & 4.070 & -2.710 &   1\\   
6265.141 & 2.180 & -2.510 &   2 &      6750.164 & 2.420 & -2.580 &   2 \\     
6270.231 & 2.860 & -2.550 &   2 &      6753.465 & 4.560 & -2.350 &   2 \\     
6280.622 & 0.860 & -4.340 &   2 &      6756.547 & 4.290 & -2.780 &   1\\   
6290.548 & 2.590 & -4.360 &   2 &      6786.860 & 4.190 & -1.900 &   2 \\     
6297.799 & 2.220 & -2.700 &   2 &      6793.260 & 4.070 & -2.430 &   1\\   
6301.508 & 3.650 & -0.720 &   3&       6796.120 & 4.140 & -2.400 &   1 \\  
6303.466 & 4.320 & -2.620 &   2 &      6804.297 & 4.580 & -1.850 &   2 \\     
6311.504 & 2.830 & -3.160 &   2 &      6806.856 & 2.730 & -3.140 &   2 \\     
6315.814 & 4.070 & -1.670 &	 1&    6810.267 & 4.610 & -1.000 &   2 \\     
6322.694 & 2.590 & -2.380 &   2 &       Fe~{\sc ii} & & \\		      
6330.852 & 4.730 & -1.220 &   2 &      5525.135 & 3.270 & -4.040 &   2 \\     
6335.337 & 2.200 & -2.280 &   2 &      5534.848 & 3.240 & -2.750 &   2 \\     
6380.750 & 4.190 & -1.340 &	 1&    5627.502 & 3.390 & -4.140 &   2 \\     
6392.538 & 2.280 & -3.970 &   2 &      5991.378 & 3.150 & -3.550 &   2 \\     
6393.612 & 2.430 & -1.430 &   2 &      6084.105 & 3.200 & -3.800 &   2 \\     
6400.323 & 3.600 & -0.230 &	 1&    6113.329 & 3.210 & -4.120 &   2 \\     
6411.108 & 4.730 & -2.330 &	 1 &   6149.249 & 3.890 & -2.720 &   2 \\     
6411.658 & 3.650 & -0.600 &   2 &      6239.948 & 3.890 & -3.440 &   2 \\     
6421.360 & 2.280 & -1.980 &   2 &      6247.562 & 3.870 & -2.320 &   2 \\     
6436.411 & 4.190 & -2.400 &	 1&    6369.463 & 2.890 & -4.210 &   2 \\     
6481.878 & 2.280 & -2.940 &   2 &      6383.715 & 5.550 & -2.090 &   2 \\     
6498.945 & 0.960 & -4.660 &   2 &      6416.928 & 3.890 & -2.700 &   2 \\     
6518.373 & 2.830 & -2.560 &   2 &      6432.683 & 2.890 & -3.580 &   2 \\     
6533.940 & 4.560 & -1.280 &   2 &      6456.391 & 3.900 & -2.100 &   2 \\     
6574.254 & 0.990 & -4.960 &	 1&    6516.083 & 2.890 & -3.380 &   2 \\   
\hline
\hline
\end{tabular}
\end{table*}

\subsection{Equivalent widths}\label{ews}

The continuum tracing and normalization of the spectra were carried out using
the task CONTINUUM within IRAF, dividing the spectra in small regions (50 \AA)
and visually checking the output. The
$EW$ measurements were carried out with the
program SPECTRE, developed by Chris Sneden (see Fitzpatrick
\& Sneden \cite{spectre}), which performs a Gaussian fitting
of the line profiles. The values are available in
electronic tables, where the first two columns
list the wavelengths and element
-- Fe~{\sc i} or Fe~{\sc ii} -- and the others show 
the corresponding $EW$ for each star.

\begin{figure*}
\psfig{figure=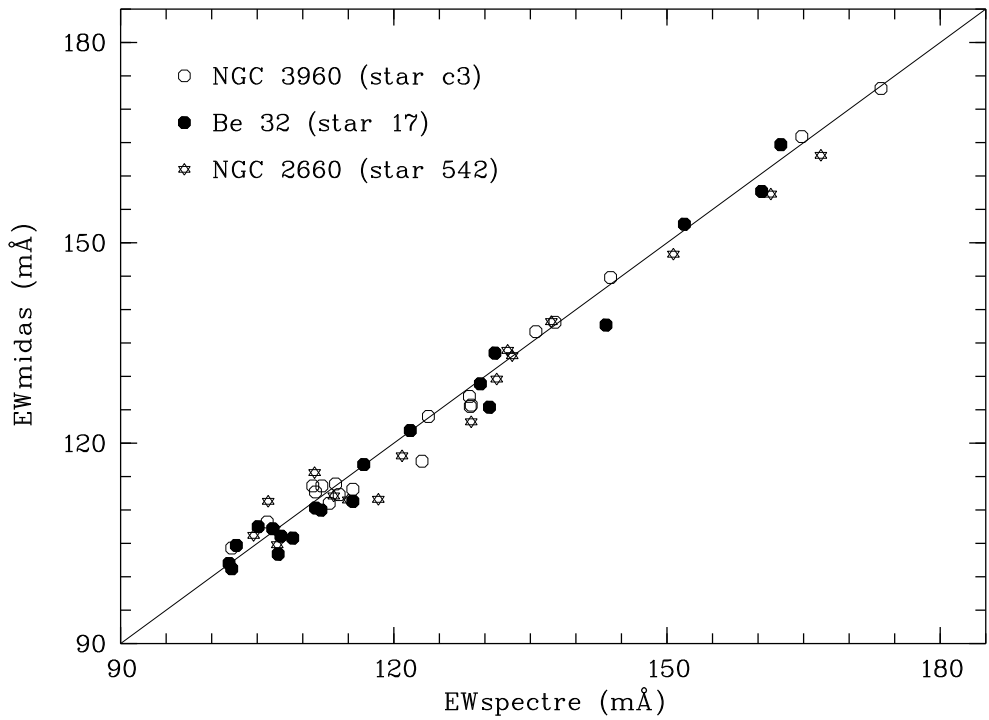, width=10cm, angle=-90}
\caption{Comparison between $EW$  measured
with SPECTRE and with MIDAS
for strong Fe~{\sc i} lines in sample stars
of the three clusters.}\label{EWstrong}
\end{figure*}

Continuum tracing and $EW$ determination are among the most  critical steps in
chemical abundance analysis, and they can represent the main reason for
discrepancies between the results obtained by different  authors (see Sect.~5).
In particular, these steps are very problematic for metal-rich giant stars,
which suffer from heavy line blending.
We also mention that Gaussian fitting of the line profile
is not always appropriate for strong lines; in those cases other fitting
techniques, such as the use of a Voigt profile or a direct integration,
would ensure that the contribution of damping wings is included
in the measurement. Since the program SPECTRE does not allow
performance of a fit different from Gaussian,
we discarded lines with $EW$s larger 
than $\sim$ 150 m\AA~(with the exception
of a couple of cases, see Fig.~\ref{EWstrong}).
On the other hand, the Gaussian function
produces a good fit of the lines with $EW$ in the range 100--150 m\AA.
This has been proved by measuring the lines with strength
in this interval also with the task
INTEGRATE/LINE within the program MIDAS, which performs
a direct integration of the spectral features.
We report in Fig.~\ref{EWstrong} a comparison between the
$EW$s measured with SPECTRE and MIDAS for three representative stars
in the clusters.
Note the very good agreement between the two sets of measurements for each
 star, indicating that the Gaussian fitting is appropriate for
features not stronger than $\sim$150 m\AA.

\subsection{Stellar parameters}\label{par}

Initial effective temperatures were derived from $B$, $V$, and $K$
photometry, applying the calibration by Alonso, Arribas, \& Martinez-Roger (\cite{alonso}),
based on a large sample of field and open-cluster giant stars.
The initial surface gravity was computed as 
$\log g=\log (M/M_{\odot})+0.4(M_{bol}-M_{bol{\odot}})+4{\cdot}\log (T_{\rm {eff}}/T_{\rm{eff}\odot})+\log g_{\odot}$, where $M$ is the mass and $M_{bol}$ 
the bolometric magnitude (with the symbol $\odot$ referring  to the Sun and 
$M_{bol{\odot}}$=4.72).
The clump masses were retrieved from the isochrones computed by the Padova
group (Bertelli et al.~\cite{padova}): 1 $M_{\odot}$ for Be~32 (age $\sim$ 6
Gyr), and 2 $M_{\odot}$ for NGC~3960 and NGC~2660  (age $\sim$ 1 Gyr). We
adopted indicative ages, but the surface gravity is only slightly affected by
the choice of clump mass: for example, assuming 1.1 $M_{\odot}$ instead of 1.0
$M_{\odot}$ would imply a change of $\sim$0.04 dex in $\log g$.

In addition, the photometric \teff~and $\log g$ only represent starting values,
whereas the two parameters were optimized during the spectral analysis. More
specifically, we employed the driver ABFIND in MOOG to compute Fe abundances
for the stars:  the final effective temperature was chosen in order to
eliminate possible trends in $\log n(\rm {Fe}$~{\sc i}) vs.~$EP$ (excitation
equilibrium). As is well known, this method relies on the circumstance that, when
the \teff~is over-estimated, the observed $EW$s of the lines with higher $EP$
are matched by a lower abundance, and vice versa.

The surface gravity was optimized by assuming the ionization equilibrium
condition, i.e.  $\log n(\rm {Fe}$~{\sc ii})$-$$\log n(\rm {Fe}$~{\sc i})=0.05
(as found for the Sun). Then, 
if necessary the \teff~was re-adjusted in order to
satisfy both the ionization and excitation equilibria.

The choice of the  microturbulence velocity deserves a more detailed
description. In most of the chemical analysis present in the literature $\xi$
is optimized by minimizing the slope of the relationship between $\log n(\rm
{Fe}$~{\sc i}) and the observed $EW$s. This technique is based on the fact that
strong lines are very sensitive to microturbulence: therefore, too high a $\xi$
would yield too low an abundance for strong lines. On the other hand, Magain
(\cite{magain}) showed that \emph{random} errors in $EW$s  lead to 
\emph{systematic} over-estimation of the microturbulence. A possible method of
obtaining 
an unbiased determination of $\xi$ is to optimize it by
instead using  the
\emph{expected} $EW$s,
which are free from random errors. The theoretical $EW$s can be computed as
$\log EW_{exp}=\log gf-EP\cdot(5040/(0.86\cdot{T_{\rm{eff}}}))$ m\AA, which is
an approximation of a classical expression (Gray \cite{gray}).

For simplicity  and for self-consistency, we adopted a relation derived using
the following procedure by Carretta et al.~(\cite{carretta04}), who 
analyzed a sample of stars in three open clusters with nearly solar metallicity
and derived a relationship between the final microturbulence velocities
(optimized using the method suggested by Magain \cite{magain}) and $\log g$.
The resulting expression is $\xi=[1.5-0.13\times{\log g}]$ km s$^{-1}$. We
tested this relationship for our sample of stars: namely, we constructed plots
of $\log n(\rm {Fe}$~{\sc i})  vs.~expected $EW$s  and found that the
microturbulence obtained by zeroing the slope of this relationship is very
similar to that 
which is computed following the prescriptions by Carretta et al. In
fact, considering the whole sample of stars in the three clusters, we obtained
a relationship between $\xi$ and $\log g$
that is  almost identical to the expression
reported above.

We also performed  a comparison with a more commonly used analysis by
optimizing the microturbulences using the $\log n(\rm {Fe}$~{\sc i})
vs.~observed $EW$s.  We find reasonable differences between the final
atmospheric parameters derived with the two techniques: in the worst cases we
have $\Delta{\xi}\sim$0.15 km s$^{-1}$, $\Delta{T_{\rm{eff}}}\sim$80 K,
$\Delta{\log g}\sim$0.3 dex, but  generally average differences in \teff,
microturbulence, and gravity do not exceed 50 K, 0.1 km s$^{-1}$, and 0.2 dex,
respectively. 
Variations in the atmospheric parameters concur in different ways to changes in
abundances, and 
the differences in the final Fe content are within $\pm$0.02 dex.
Furthermore, we notice that in some cases the relationship $\log n(\rm
{Fe}$~{\sc i}) vs.~observed $EW$ has a negligible slope, even if the
microturbulence has been optimized in function of the expected $EW$s.

\subsection{Errors}\label{errori}

Our abundance scale is directly referred to the solar $\log n(\rm{Fe})$ and the
majority of the spectral lines used for giants stars are in common with those
included in the solar line list. In this way, internal errors due to
uncertainties in the oscillator strengths should be minimized.

Random internal errors in $EW$s can be estimated
by comparing the $EW$s of two stars with similar parameters. Considering the
common
lines  between the two stars and rejecting significant outliers, we
found the following average differences: $<\Delta{EW}>\pm{\delta}=-0.95\pm2.54$
m\AA~(where $\delta$ indicates the rms)  for NGC~3960 (c3$-$c6; 112 lines),
$-1.02\pm3.75$ m\AA~for Be~32 (17$-$794; 90 lines), and $-1.49\pm4.48$ m\AA~for
NGC~2660 (318$-$862; 76 lines). The random errors in $EW$ are represented by
${\delta}/{\sqrt{2}}$ (assuming that they can be equally attributed to both
stars in the  pair under consideration), therefore we find that their values
are $\sim$1.8, $\sim$2.7, and $\sim$3.2 m\AA~for NGC~3960, Be~32 and NGC~2660,
respectively.

The effect of random errors in $EW$s and of 
errors in the atomic parameters on the
derived abundance for a single star is well-represented by $\sigma_1$, the
standard deviation from the mean abundance based on the whole set of lines (see
Table~\ref{tab_Fe}).
However, Fe abundances are 
also affected by uncertainties on the adopted
stellar parameters: the total error in [Fe/H] for each star,
$\sigma_{\rm{tot}}$, can be computed by quadratically adding $\sigma_1$ to the
error deriving from random uncertainties in \teff, $\log g$ and $\xi$, which we
will call $\sigma_2$. Typical $\sigma_2$ values for each cluster can be
estimated by varying one parameter at a time (holding the others fixed) and
then by quadratically adding the three related errors, 
$\sigma_{T\rm{eff-rd}}$, $\sigma_{\log g-\rm{rd}}$, and $\sigma_{\xi-\rm{rd}}$
(see below).

Errors in \teff, $\log g$, and $\xi$ have been evaluated in the same
fashion as in Carretta et al.~(\cite{carretta04}).  A detailed description of
the method can be found in the quoted reference, but
 we mention here that  we find
standard total  internal errors of $\sim$35 K in temperature for
 individual stars in the three clusters. The
corresponding  uncertainty in abundance ($\sigma_{T\rm{eff-TOT}}$) is the 
quadratic sum of the  internal
random term ($\sigma_{T\rm{eff-rd}}$, related to the $EW$
measurement) and the  internal
systematic term ($\sigma_{T\rm{eff-sys}}$).  Using the
formulas of Carretta et al., we estimated $\sigma_{T\rm{eff-rd}}$ to be 0.033,
0.055, and 0.066 dex for NGC~3960, NGC~2660, and Be~32, respectively,  i.e. about
47 \%, 69 \% ,and 78 \% of the total uncertainty. Therefore, the internal random
errors in the temperatures are $\sim$15--25 K
for individual stars in these clusters.
 Systematic scale errors are always difficult to estimate, and a more
detailed discussion must be deferred to the completion
of the analysis for our whole sample of clusters. At present, the good
agreement between results from different approaches (photometric
and spectroscopic) allows us to
estimate that scale errors are likely to be confined typically
within 100--150 K.

Errors in surface gravities are due to two contributions, one from
spectroscopic effective temperatures,  and the other  from errors in the
measurement of individual lines (see Carretta et al.~\cite{carretta04}). The
total (random + systematic) error $\Delta{\log g}$ results in 0.15--0.25
dex.  Taking only the random part (i.e. 47, 69, and 78 \% for
NGC~3960, NGC~2660, and Be~32, respectively) of this error
into account, we find that the
total random uncertainty $\Delta{\log g}_{\rm{rd}}$ is 0.13 dex for NGC~3960,
0.10 dex for NGC~2660 and 0.12 dex for Be~32.

Errors in the microturbulence velocity are computed for a typical star in each
cluster and depend on the variation in the slope of the adopted abundance
vs.~expected $EW$ relation and on the relationship between $\xi$ and $\log g$
(see Carretta et  al.~\cite{carretta04}). The random internal errors
$\Delta{\xi}_{\rm {rd}}$ are $\sim$0.10 km s$^{-1}$ for NGC~3960, $\sim$0.17 km
s$^{-1}$ for NGC~2660 and $\sim$0.20 km s$^{-1}$ for Be~32.

As already mentioned, $\sigma_2$ is the error in abundance related to random
internal uncertainties in the stellar parameters, and it is due to three terms:
$\sigma_{T\rm{eff-rd}}$, $\sigma_{\log g-{\rm rd}}$, and $\sigma_{\xi-{\rm
rd}}$. The first term is reported above, while the two last
are calculated by
estimating the sensitivity of $\log n(\rm{Fe}$~{\sc i}) to the following
changes: $\Delta{\log g}_{\rm{rd}}$=0.10--0.13 dex and $\Delta{\xi}_{\rm
{rd}}$=0.10--0.20 km s$^{-1}$. The sensitivity of [Fe/H] to errors in the
stellar parameters is exemplified in Table~\ref{sens}: for each cluster we
chose the star with Fe abundance and \teff \ most similar to the average
values.
Note that we adopted sample parameter variations of $\pm$100 K in
\teff, $\pm$0.2 dex in $\log g$ and $\pm$0.15 km s$^{-1}$ in $\xi$, even if
different from the random errors in the stellar parameters estimated to compute
$\sigma_2$ in our clusters.

The presence of systematic errors due to the method of analysis can be checked
by analyzing stars with well-known metallicity, as for example the
\object{Hyades}, and possibly observed with the same
instrument. Unfortunately,
we observed two Hyades stars (the clump objects $\gamma$ Tau and
$\delta$ Tau) only with SARG at the TNG at
lower resolution ($R\sim29,000$). The same method described
above was employed to carry out the analysis.
We found similar abundances and  parameters for the two stars,
with \teff=4860 K,
$\xi$=1.40 km s$^{-1}$, $\log g$=2.50, and [Fe/H]=+0.17$\pm$0.08
for $\gamma$ Tau, and \teff=4905 K,
$\xi$=1.41 km s$^{-1}$, $\log g$=2.60, and [Fe/H]=+0.19$\pm$0.08
for $\delta$ Tau. Then
$\xi$ was optimized using both
the observed and expected $EW$. We note that using the parameters
reported above we were indeed able to optimize the Fe abundance vs. $EW$s distribution
for either expected and observed $EW$s.

Our value for the metallicity of the Hyades appears somewhat
higher than most of the estimates present in the literature based on dwarfs:
for example Boesgaard \& Friel (\cite{bf90}) quoted [Fe/H]=+0.13 for MS stars,
which is the most commonly accepted value.  However, considering that they
assume $\log n(\rm{Fe})_{\odot}$=7.56, higher by 0.07 than our value, our
metallicity is perfectly consistent with theirs. In more recent analysis
Boesgaard, Beard, \& King (\cite{bbk}) quote [Fe/H]=+0.16,
while Paulson, Sneden, \& Cochran (\cite{paulson}) find
[Fe/H]=+0.13. Recent investigations
of giant stars are those by Smith (\cite{smith}:
[Fe/H]=+0.15), by Wylie, Cottrell, \& Taute (\cite{wylie}: [Fe/H]=+0.19), and
by Schuler et al.~(\cite{schuler}: [Fe/H]=+0.16), in agreement with our result.

\section{Results}\label {risultati}

Our results for the Fe abundances are reported in Table~\ref{tab_Fe}. 
Columns 2
and 3 list the photometric \teff~derived from $(B-V)$ and $(V-K)$ colors,
and
the average of these two values
was assumed as initial \teff~for the analysis for each star.
In Col.~4 we report the photometric $\log g$ (mean value
of gravities from
$B-V$ and $V-K$ colors). For star 938 in Be~32
$BV$ photometry by D'Orazi et al.~(\cite{dorazi}) was not available; therefore,
we adopted the $V$ magnitude by Kaluzny \& Mazur (\cite{kal}) and
listed only \teff$(V-K)$. For star 296 in NGC~2660, the $K$ value is clearly
wrong, leading to a \teff($V-K$) of 4300 K, very different from the
\teff($B-V$) and  from the temperatures of the other clump stars. For
the whole sample, effective temperatures derived from ($B-V$) and ($V-K$)
colors are in generally good agreement (i.e. within the errors), differing by
0--70 K in NGC~3960, by 20--150 K in Be~32, and by $\sim$100 K in NGC~2660.

The final stellar parameters determined via the spectroscopic analysis
described in Sect.~\ref{par} are shown in  Cols 5--7 (\teff, $\log g$ and
$\xi$). In most cases, we found  reasonable agreement between the photometric
and spectroscopic parameters. The spectroscopic
and (average) photometric temperatures usually coincide within $\sim$100 K.

As far as surface gravities are concerned, we note that  to satisfy the
ionization equilibrium, we had to consider a $\log g$ usually lower than the
photometric ones. The differences have average values of 
$\Delta({\log g}_{\rm{phot}}-{\log g}_{\rm{spec}})$=0.30$\pm$0.12 (statistical
error) for NGC~3960, 0.20$\pm$0.10 for Be~32, and 0.15$\pm$0.12 for NGC~2660.
These differences  are slightly larger than the errors in $\log g$ (see
Sect.~\ref{errori}) and could be attributed to several random  factors, such
as internal errors, errors in distance moduli (0.2 mag translates into 0.08 dex
in gravity), errors in reddenings (important especially for NGC~3960 where a
differential reddening was noted by Prisinzano et al. 2004 and confirmed for
the inner region by B06a at the level of $\pm$0.05 mag, corresponding to
$\Delta{\log g}\sim0.07$ dex), errors  in ages (i.e., in masses, although the
contribution is rather small), non homogeneity of the data sources,  etc.
 Another possible explanation for the difference
between spectroscopic and photometric gravities might
be represented by non-LTE effects and/or inadequacies in classical (1-d) model
atmospheres where important features such as spots,
granulation, activity, etc.~are neglected.
Such a discrepancy has also been
encountered by other authors (Feltzing \& Gustafsson \cite{feltzing};
Schuler et al. \cite{schuler03}; 
Allende Prieto et al.~\cite{allende04}) in studies
of cool metal-rich stars.
Nevertheless, random uncertainties might not be the only reason for the
discrepancies between photometric and spectroscopic gravities,
as also a
systematic error due to the method of analysis could probably be present.
The microturbulence values, derived as described in Sect.~\ref{par}, cover a
rather narrow range in each cluster,  as expected from the fact that clump
stars are in the same evolutionary status.

In the last three columns of Table~\ref{tab_Fe}, 
we report the Fe abundance with
respect to the Sun ([Fe/H]), 
and their errors: $\sigma_1$ and $\sigma_{\rm{tot}}$.
$\sigma_{\rm{tot}}$ are computed by quadratically adding $\sigma_1$ and
$\sigma_2$ as discussed in Sect.~\ref{errori}.  We estimated typical $\sigma_2$
values in the three clusters of $\sim$0.05 dex for NGC~3960, and 0.08--0.09 dex
for the other two clusters.
When determining Fe abundances of the stars,
1$\sigma$-clipping was performed as the first step, 
that is, before the optimization of the stellar
parameters.

The average (weighted mean) values for the clusters are also given, with the
standard deviation from the mean (considering the total uncertainty
$\sigma_{\rm{tot}}$). NGC~3960 turns out to have a solar metallicity,
$<$[Fe/H]$>$=$+0.02\pm0.04$ dex, at variance with some previous  reports of
sub-solar Fe content (see the discussion in Sect.~\ref{disc1}). NGC~2660
also
has a solar Fe content: $<$[Fe/H]$>$=$+0.04\pm0.04$, while for Be~32 we derive
a mean value $<$[Fe/H]$>$=$-0.29\pm0.04$ in fair agreement with previous
estimates of sub-solar metallicity.

\begin{figure*}
\psfig{figure=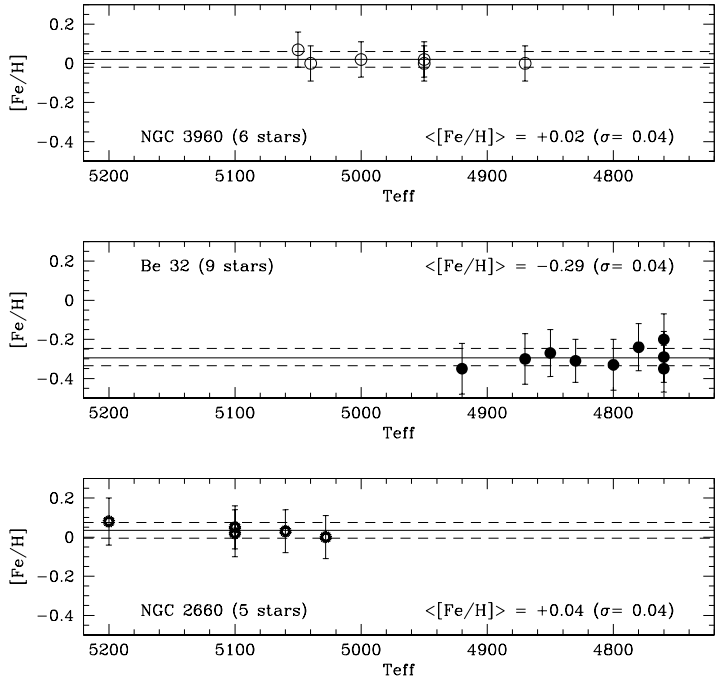, width=10cm, angle=0}
\caption{Fe abundances ([Fe/H]) as a function of effective temperature for
stars in the three clusters. From top to bottom: NGC~3960, Be~32 and NGC~2660.
Error bars ($\sigma_{\rm{tot}}$, see Sect.~\ref{risultati}) are also shown. The
solid lines represent the average [Fe/H] (weighted mean) for the cluster, while
the dashed ones indicate the standard deviation from the mean
(rms).}\label{ferro}
\end{figure*}

Plots of the Fe abundance as a function of \teff~are shown in Fig.~\ref{ferro}
for each cluster. Error bars ($\sigma_{\rm{tot}}$) are also reported.
The resulting abundances
are characterized by a 
small dispersion. This is only in part due to the 
1-$\sigma$ clipping performed during the first iteration
of the analysis; indeed, most of the lines discarded during that step
were common to the various stars. 
Typical $\sigma_1$ values before the clipping were
about 0.15 dex (i.e. twice the final $\sigma_1$).

\begin{table*}[!] \footnotesize
\caption{Stellar parameters (photometric
and spectroscopic) and Fe abundances
for the sample stars in the three clusters.}\label{tab_Fe}
\begin{tabular}{llllllllllllll}
\hline
\hline
Star &\teff$_{(B-V)}$ &\teff$_{(V-K)}$ & $\log g_{\rm{phot}}$* & \teff$_{\rm{spec}}$ &
$\log g_{\rm{spec}}$ & $\xi$ & [Fe/H] &$\sigma_{1}$ & $\sigma_{\rm{tot}}$  \\
 &(K) &(K) & & (K) & & km s$^{-1}$ & & &\\
\hline
NGC~3960 &&&&&& &\\
\hline
c3 &4814 &4743 &2.78 &4950 &2.35 &1.19& +0.00 &0.08 & 0.09\\
c4 &5047 &5027  &2.78 &5050 &2.54 &1.17& +0.07 & 0.07& 0.09\\
c5 &4862 &4769  &2.62 &4870 &2.16 &1.22& +0.00 &0.07 & 0.09\\
c6 &5120 &5121  &2.72 &4950 &2.4  &1.19& +0.02 & 0.08 &0.09\\
c8 &5064 &5017  &2.73 &5040 &2.57 &1.18& +0.00 &0.07  &0.09\\
c9 &4968 &4959  &2.71 &5000 & 2.45&1.18 & +0.02& 0.08 &0.09\\
Average Fe & & & & & & & +0.02&  &0.04 (rms)\\
\hline
Be~32  & &&&&&& & \\
\hline
17 &4830 &4738  & 2.45&4830 &2.22 &1.21 &$-$0.31 &0.07 & 0.11\\
18 &4866 & 4784 & 2.47&4850 &2.27 &1.21 &$-$0.27 &0.08 & 0.12 \\
19 &4822 &4694  & 2.45&4760 &2.26 &1.21 &$-$0.35 &0.08 & 0.12\\
25 &4718 & 4598 & 2.61&4760 &2.40 &1.19 &$-$0.20 &0.10 & 0.13 \\
27 & 4729& 4622 & 2.67&4780 &2.35 &1.19 &$-$0.24 &0.08 & 0.12 \\
45 & 4964& 4905 & 3.15&4920 &3.00 &1.11 &$-$0.35 &0.10 & 0.13 \\
938 &-- &4732& 2.45&4870 &2.33 &1.19 &$-$0.30 &0.09 & 0.13 \\
940 &4800 &4710 & 2.45&4800 &2.10 &1.23 &$-$0.33 &0.09 & 0.13\\
941 &4793 &4690 & 2.43&4760 &2.40 &1.19 &$-$0.29 &0.08 & 0.13\\
Average Fe&&& & & & & $-$0.29& & 0.04 (rms)\\
\hline
NGC~2660 &  &&&&&& & \\
\hline
296 &5126 & --&3.01 &5200 &3.01 &1.11 &+0.08 &0.09 & 0.12\\
318  &5028 &4926  &2.75 &5030 &2.59 &1.16 &0.00  &0.07 & 0.11\\
542  & 5008&4925  &2.76 &5060 &2.48 &1.17 &+0.03 &0.07 & 0.11\\
694 &4996 &4886  &2.85 &5100 &2.77 &1.14 &+0.05 &0.08 & 0.11\\
862  &5036 &4940  &2.85 &5100 &2.60 &1.16 &+0.02 &0.09 & 0.12\\
Average Fe  && & & & & & +0.04& & 0.04 (rms)\\
\hline
\hline
\end{tabular}

*Photometric gravities are the average between $\log g$ from $(B-V)$ and $(V-K)$.
\end{table*}
\begin{table*}[!] \footnotesize
\caption{Sensitivities of Fe abundances to variations in 
the atmospheric parameters for typical stars in the three clusters.}\label{sens}
\begin{tabular}{llllllllllllll}
\hline
\hline
Star &$\sigma_{T\rm{eff}}$ &$\sigma_{\log g}$ &$\sigma_{\xi}$  \\
&$\Delta$\teff=$\pm$100 K & $\Delta{\log g}$=$\pm$0.2 dex & $\Delta{\xi}$=$\pm$0.15 km s$^{-1}$\\
\hline 
c9 (NGC~3960) & +0.09/$-$0.06&+0.01/0.00 &$-$0.05/+0.06 \\
17 (Be~32) &+0.09/$-$0.07&+0.01/0.00&$-$0.05/+0.06\\
542 (NGC~2660) &+0.09/$-$0.07&+0.01/0.00&$-$0.05/+0.06\\
\hline
\hline
\end{tabular}
\end{table*}

\section{Discussion}\label{disc}

\subsection{Comparison with previous studies}\label{disc1}

Using the synthetic CMD
technique, Sandrelli et al.~(\cite{sandrelli}) found
for NGC~2660 that all  the best-fit models for the  three adopted sets of
stellar evolutionary tracks were obtained for solar metallicity. For Be~32,
Friel et al.~(\cite{friel02}) derived [Fe/H]=$-$0.50 from low-resolution
spectra. Sub-solar metallicity  has also been found by Kaluzny \& Mazur
(\cite{kal}) on the basis of photometry ([Fe/H]=$-$0.37$\pm$0.05) and by
D'Orazi et al.~(\cite{dorazi}) who derived a best-fit metallicity $Z=0.008$
([Fe/H]$\sim-$0.40). While our results are reasonably consistent with the
previous ones,  it is important to emphasize that we provide the first
metallicity reports for these two open clusters based on high-resolution
spectroscopic data.

As far as NGC~3960 is concerned, literature reports based on photometry or low
resolution spectroscopy quoted a clearly sub-solar Fe content (Friel \& Janes
\cite{fj93}; Geisler, Clari\`a, \& Minniti \cite{geisler}; Piatti, Clari\`a, \&
Abbadi \cite{piatti}). The discrepancy with these results is probably due to
the difference in the reliability of the method of analysis and quality of the 
data, so we think it is not worth undertaking a detailed comparison.
On the other hand, 
we can  perform a direct comparison with  the very recent study by B06a, who
carried out a spectroscopic analysis (Fe and other elements) for three stars in
the cluster, observed with the FEROS spectrograph at the 1.5m ESO telescope in
La Silla (Chile). The three objects in common with our sample are c5, c6 and
c8. While the $S/N$ ratio of their spectra is lower than ours, the spectral
resolving power of the two instruments is comparable.

The same line list (with the same $\log gf$ and damping parameters) was used to
determine the Fe abundances; furthermore, the same concepts for optimizing the
stellar parameters were adopted. Nevertheless, the two methods of  analysis
differ in other  important steps: the determination of the continuum and $EW$
measurement and the adopted spectral code (the package ROSA, developed by
Gratton \cite{ROSA} has been used by B06a). Detailed descriptions of the method
of analysis adopted by B06a can be found in Bragaglia et
al.~(\cite{bragaglia01}), G03, and Carretta et al.~(\cite{carretta04}).

The comparison between our results and those by B06a is presented in
Table~\ref{confr}, where a very good agreement between the spectroscopic parameters
derived by the two methods is evident. In the last column
we list the difference
among the net $\log n(\rm{Fe})$ values from the two studies. Note
that B06a have $\log n(\rm {Fe}$~{\sc i})$_{\odot}$=7.54 and $\log n(\rm
{Fe}$~{\sc ii})$_{\odot}$=7.49 for the Sun (while we find Fe~{\sc ii} to be
higher than Fe~{\sc i}, see Sect.~\ref{sun}). We carefully checked the reasons
for this difference between the solar Fe content derived with MOOG and ROSA: in
the case of neutral Fe, the abundance obtained with ROSA for the Sun is
systematically higher,  for both strong and weak lines. Had we found higher
differences in the case of strong lines, this could have been related to a
faulty 
treatment of damping. The difference between the abundances of ionized Fe
goes instead in the opposite direction with respect to Fe~{\sc i}. We therefore
tentatively conclude that the discrepancy of $\pm$0.05 dex among the two solar
analyses could be due to a systematic difference between the two combinations
of spectroscopic code and model atmospheres.
Considering this, the average difference between the metallicity obtained by us
and by B06a for the three stars in common is 0.07 dex (our estimate minus the
B06a one).

By comparing the $EW$ sets for each star, we found that our measurements are
systematically higher than those of B06a, with average differences of
$\sim$2--5 m\AA~($\sigma{\sim}\pm6$ m\AA). The reason for such a discrepancy
should be attributed to the determination of the continuum and the method of
measurement. In order to check this hypothesis, we performed a complete
re-analysis (i.e. starting from the normalization of the spectrum) of the FEROS
spectrum of star c6 used by B06a, for which the largest difference
$\Delta$[Fe/H]$_{\rm{MOOG-ROSA}}$ is observed. We considered only the portion
of the spectrum in common with the UVES range and found the FEROS and UVES
$EW$ sets to be nearly identical  (average difference $\sim$1.5 m\AA, instead
of $\sim$3 m\AA~when we use the original B06a $EW$s). This confirms that one of
the sources of systematic errors could be continuum determination
and measurement of $EW$s.
In addition, as mentioned above, systematic differences due to the adopted
spectral analysis package are present.
Indeed, we carried out the analysis of star c6
with MOOG using the $EW$s as measured by B06a and  find  a result almost
identical to ours, that is, [Fe/H]=+0.02, \teff=5000 K, $\log g$=2.5, $\xi$=1.18
km s$^{-1}$. All these issues, together with random differences in the final
stellar parameters can explain the offset between the two analyses. The average
difference (+0.07 dex) is compatible within the error bars. Note that, although
the metallicities of  B06 are in the three cases lower than ours,  this is not
a completely systematic offset. A random contribution is 
also present,
reflected in the fact that the differences between our results
and B06a's
change from one star to the other.

In Sect.~\ref{risultati} we mentioned the good agreement between the
photometric \teff, assumed as initial parameters, and the final \teff~derived
during the spectroscopic analysis. The only star for which we found a rather
significant discrepancy is c6, where
 the \teff~optimized during the analysis is 170 K
cooler than the initial one. We think we can safely assume that the
spectroscopic temperature is the right one for two reasons: first, our value is
in good agreement with that of B06a. Second, we compared the spectrum of star
c6 with those of the other clump objects with similar parameters, and the
spectral features confirmed the results. The reason for the wrong photometry of
star c6 can be ascribed to the fact that the cluster is affected by
differential reddening, as found by Prisinzano et al.~(\cite{prisinzano}) and
B06a. That the photometry leads to a hotter \teff~than the real one (which we
assume to be the spectroscopic one) could imply that star c6 is not
affected by $E(B-V)$=0.29, but by a lower reddening.  This agrees with
the findings of B06a, which re-derived the reddening for the three stars in
their sample from the spectroscopic \teff. In this way, star c6 turned out to
have  $E(B-V)$=0.22: assuming a reddening difference of 0.07 dex  justifies an
error of $\sim$150 K in \teff.

\begin{table*}[!] \footnotesize
\caption{NGC~3960: comparison with the results by Bragaglia
et al.~(\cite{bragaglia06}); numbers
in parenthesis are their IDs.}\label{confr}
\begin{tabular}{lllllll}
\hline
\hline
star & & T$_{\rm{eff}}$ & log g & $\xi$ & [Fe/H] & $\Delta{\log n(\rm{Fe})}$ \\
 & & K & & km/sec \\
\hline
&&&&& \\
c5     &MOOG & 4870 &2.16 & 1.22 &+0.00   &$-$0.09 \\
(41)   &ROSA & 4850 &2.20 & 1.21 &$-$0.14 &        \\
&&&&& \\
c6     &MOOG & 4950 &2.40 & 1.19 &+0.02   &$-$0.12 \\
(28)   &ROSA & 4900 &2.06 & 1.23 &$-$0.15 &        \\
&&&&& \\
c8     &MOOG & 5040 &2.57 & 1.17 &+0.00   &$-$0.01 \\
(50)   &ROSA & 5000 &2.70 & 1.15 &$-$0.06 &        \\
\hline
\hline
\end{tabular}
\end{table*}

\subsection{Metallicity distribution in the disk}\label{disc2}

\begin{figure*}
\psfig{figure=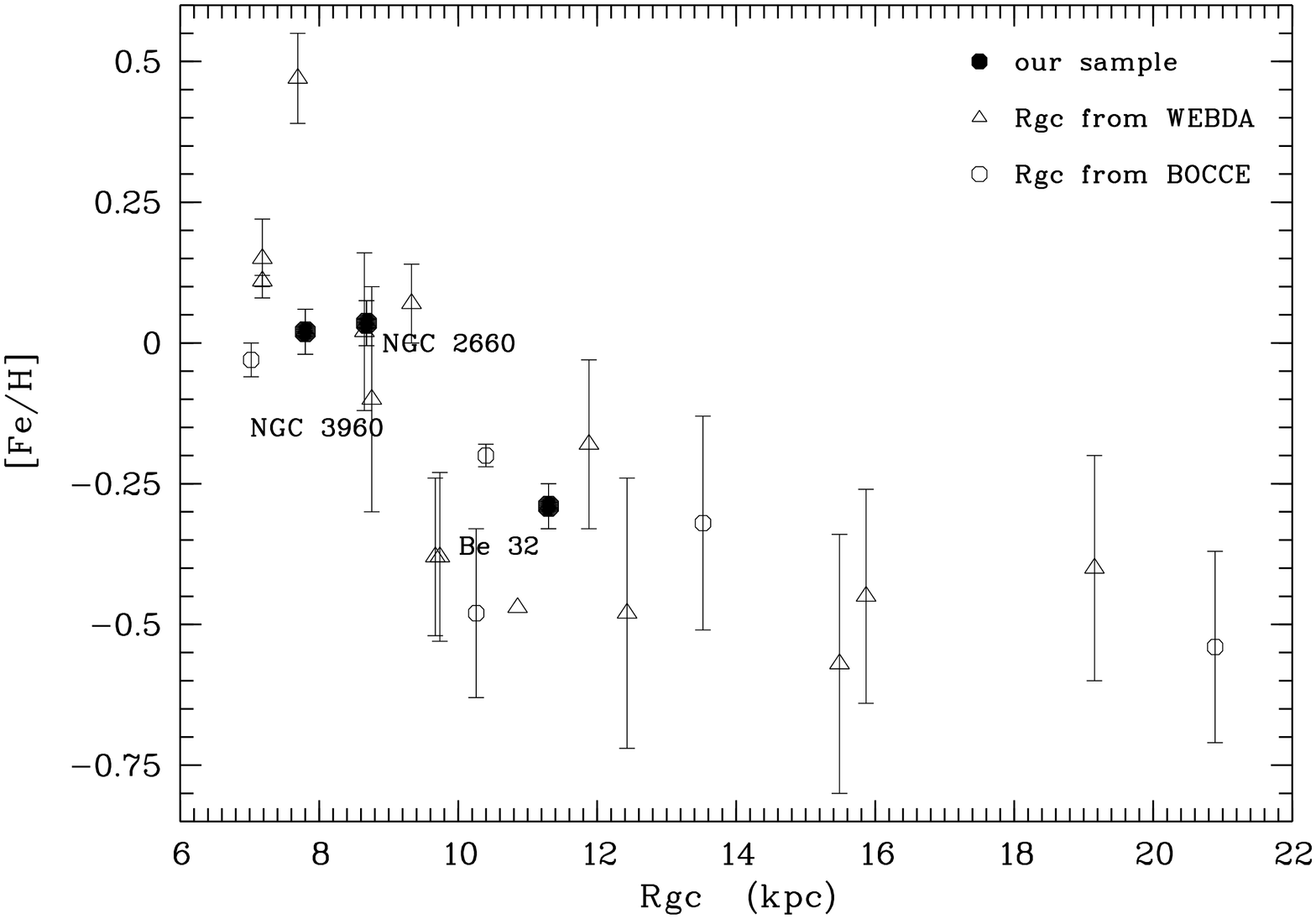, width=7cm, angle=-90}

\psfig{figure=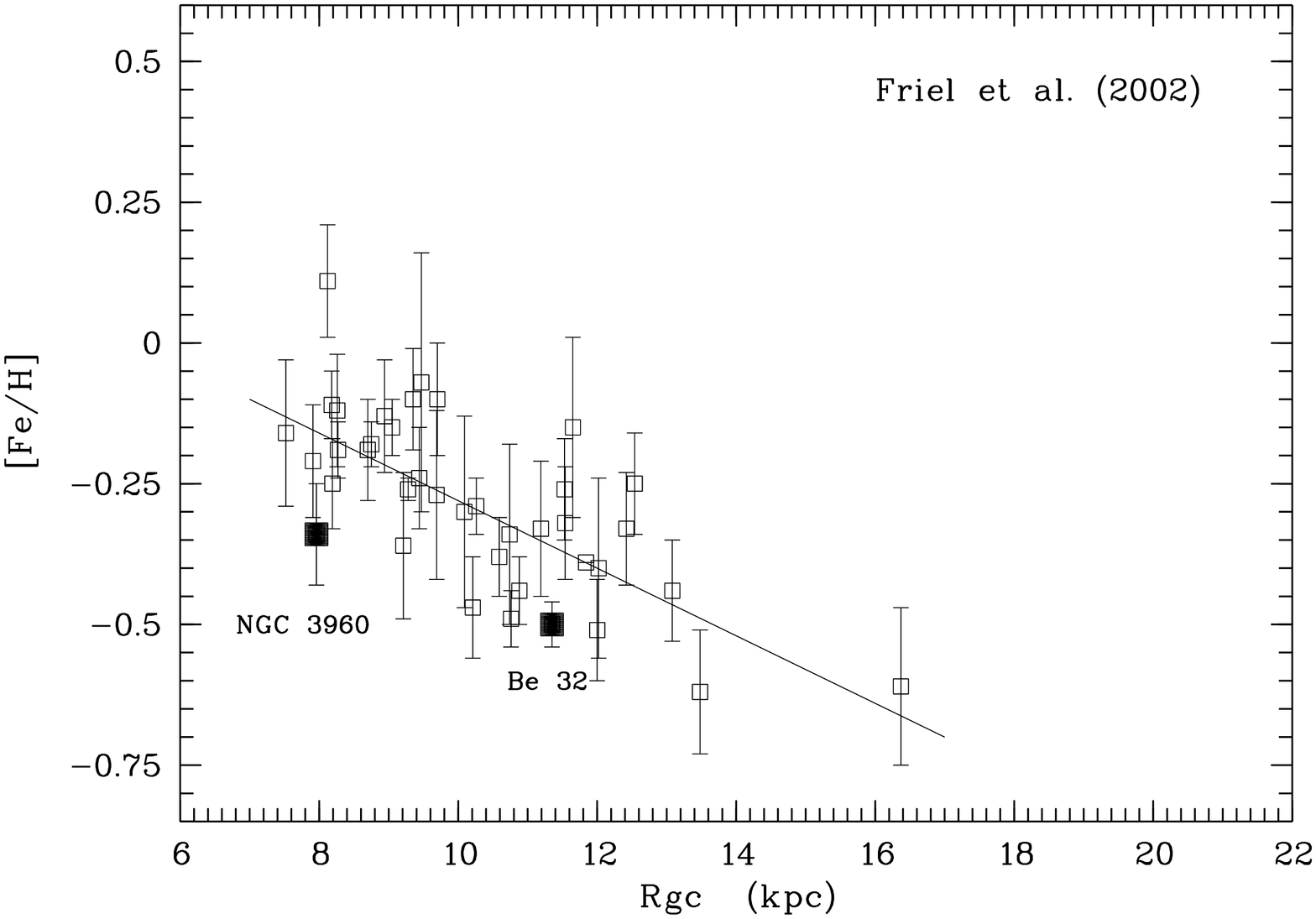, width=7cm, angle=-90}
\caption{Radial gradient ([Fe/H] vs.~Galactocentric distance) for open
clusters. In the upper panel  our results (filled circles) are compared to
other clusters analyzed with high-resolution spectroscopy (open circles and
triangles). In the lower panel  the sample analyzed by Friel et
al.~(\cite{friel02}) is shown (filled symbols are clusters in common with the
present work).}\label{grad}
\end{figure*}

How do our results contribute to the picture of
the overall metallicity distribution with
Galactocentric distance? As already mentioned in Sect.~1,  a proper comparison
should be carried out  taking into account the clusters analyzed in a homogeneous
way. However, large samples of clusters observed with high-resolution
spectroscopy and analyzed with the same method are not available yet in the
literature; therefore, we report in Fig.~\ref{grad}  plots of [Fe/H] as a
function Galactocentric distance based on two different datasets. In the upper
panel we show our data and the results from other high-resolution
studies in the literature\footnote{Part of these data are collected in Table~7 of Friel
et al.~(\cite{friel03}, see also references therein); 
others are from Carretta et
al.~(\cite{carretta04}, \cite{carretta05}),  Carraro et al.~(\cite{carraro04}),
Villanova et al.~(\cite{villanova}), Yong et al.~(\cite{yong05}), Randich et
al.~(\cite{R06M67}), Gratton et al.~(\cite{6791}); 
note that the quoted investigations are all based on
giant stars with the exception of Randich et al.~(\cite{R06M67}) 
in which dwarfs
and slightly evolved stars of \object{M~67} were considered.}. Clusters
represented by open circles are included in the BOCCE program (Bragaglia
\& Tosi~\cite{bt06}),  so we assumed the
ages and $R_{\rm{gc}}$ listed there. For
the three clusters investigated in this work, we assumed the $R_{\rm{gc}}$ and
ages reported in Table~\ref{targets}. For the remaining clusters we instead
assumed  ages and $R_{\rm{gc}}$ from the WEBDA
database\footnote{http://www.univie.ac.at/webda/}.  In the lower panel we
considered the dataset by Friel et al.~(\cite{friel02}) based on low
resolution spectroscopy, which represents a homogeneous sample, as far as the
metallicity scale is concerned. Filled symbols are clusters in common with us.

The comparison between the two panels shows  fair agreement between the
average trends of [Fe/H] with $R_{\rm{gc}}$ derived from high and low
resolution spectra, at least up to distances of $\sim$13 kpc (but see below).
Our data are consistent with the existence of a radial gradient, since the
clusters at a lower $R_{\rm{gc}}$ have higher Fe content. Notice that the
spread in the [Fe/H] vs. $R_{\rm{gc}}$ distribution \emph{at low Galactocentric
distances} appears reduced when using high-resolution data compared the
low-resolution ones (excluding \object{NGC~6791} which has [Fe/H]=+0.47). For
clusters in the high-resolution sample with  $R_{\rm{gc}}$ smaller  than
$\sim$9--10 kpc, we find that none has a sub-solar metallicity and that the slope
of the  [Fe/H] vs. Galactocentric radius distribution
in the inner 10 kpc is slightly steeper than
that derived by Friel et al.~at low resolution.

For clusters with $R_{\rm{gc}}$ larger than $\sim$13 kpc, the Friel dataset
does not allow us to distinguish whether the gradient maintains the same slope
given by the inner clusters, or it flattens with increasing distance. As
previously reported by other authors, the high-resolution data seem to
indicate instead that the gradient flattens at large Galactocentric distances.

However, we stress again that this is only an indicative comparison aimed at
understanding how our results can be inserted in the average [Fe/H] distribution in
the Galactic disk.

\section{Summary and conclusions}

We report on the Fe abundance for three old open clusters, NGC~3960, NGC~2660,
and Be~32, based on FLAMES/UVES observations of clump stars.

The data were collected within a VLT/FLAMES program on open clusters and
the primary aim was the investigation of the radial metallicity gradient in
the Galactic disk, which represents a critical constraint for models of
Galactic chemical evolution. Abundances of other elements derived from UVES
spectra and the
abundances of lithium from the GIRAFFE data of MS stars will be
presented in forthcoming papers.

In this paper we accurately describe the method of spectroscopic analysis
that will be used for all the clusters in the sample in order to build a
homogeneous database with all the abundances and effective temperatures on the
same scale. We present
the results for the first three clusters analyzed. For
the younger clusters NGC~3960 and NGC~2660 (ages $\sim$1 Gyr), we find
[Fe/H]=+0.02$\pm$0.04 (weighted average, rms) and [Fe/H]=+0.04$\pm$0.04,
respectively, while the $\sim$6--7 Gyr old Be~32 turns out to have
[Fe/H]=$-$0.29$\pm$0.04. We stress that our study represents the first high
resolution  spectroscopy metallicity investigation for NGC~2660 and Be~32. 

NGC~3960 has been
recently investigated by Bragaglia et al.~(\cite{bragaglia06}), who found
[Fe/H]=$-$0.12 (i.e. a slightly lower value than ours) from high-resolution
FEROS data. Previous reports suggested instead a clearly sub-solar [Fe/H], at
variance with us.

Finally, we discuss our findings in the context of the overall metallicity
distribution with Galactocentric radius. With the \emph{caveat} that the
comparison with other (high- and low-resolution spectroscopic) results is biased
by the fact that different methods of analysis are used by the various authors,
we tentatively conclude that our results support the presence of a negative
radial [Fe/H] gradient. In addition,  the spread in the [Fe/H]
vs.~$R_{\rm{gc}}$ distribution appears reduced when using high-resolution data
and compared to low-resolution ones. These results should be confirmed based on
a larger sample of data analyzed with a homogeneous method.

\begin{acknowledgements}
P.S. acknowledges support by the Italian MIUR, under PRIN 2003029437, and
of the Bologna Observatory, where this work was completed.
We thank the anonymous referee for his/her competent and valuable comments
and suggestions to improve the paper.
We warmly thank
C. Sneden for having provided a new version of MOOG
with updates regarding damping parameters. We are grateful to R. Gratton
for helpful discussions, and we acknowledge the use of the
WEBDA database created by J.-C.~Mermilliod.
\end{acknowledgements}
{}

\end{document}